\documentclass[onefignum,onetabnum]{siamonline220329}

\usepackage{lipsum}
\usepackage{amsfonts}
\usepackage{graphicx}
\usepackage{epstopdf}
\usepackage{algorithmic}
\usepackage[caption = false]{subfig}
\usepackage{amssymb}
\usepackage{bbm}
\usepackage{nicefrac}
\usepackage{verbatim}
\ifpdf
  \DeclareGraphicsExtensions{.eps,.pdf,.png,.jpg}
\else
  \DeclareGraphicsExtensions{.eps}
\fi

\usepackage{enumitem}
\setlist[enumerate]{leftmargin=.5in}
\setlist[itemize]{leftmargin=.5in}

\newsiamremark{remark}{Remark}
\newsiamremark{hypothesis}{Hypothesis}
\crefname{hypothesis}{Hypothesis}{Hypotheses}
\newsiamthm{claim}{Claim}

\headers{An AME Formulation of the Watts Threshold Model on Hypergraphs}{L. A. Keating, K.-I. Goh, and M. A. Porter}

\title{An Approximate-Master-Equation Formulation of the Watts Threshold Model on Hypergraphs\thanks{Submitted to the editors \today.
}}

\author{Leah A.~Keating\thanks{Department of Mathematics, University of California, Los Angeles, CA 90095, USA; Vermont Complex Systems Institute, University of Vermont, Burlington, VT 05405 USA; Department of Computer Science, University of Vermont, Burlington, VT 05405 USA; {School of Mathematical Sciences, Dublin City University, D09 K20V Dublin, Ireland. (\email{leah.keating@dcu.ie}{)}}}
\and K.-I. Goh\thanks{Department of Physics, Korea University, Seoul, South Korea. 
  (\email{kgoh@korea.ac.kr}{)}}
\and Mason A.~Porter\thanks{Department of Mathematics, University of California, Los Angeles, CA 90095 {USA}; Department of Sociology, University of California, Los Angeles, CA 90095 {USA}; 
Nordita (Nordic Institute for Theoretical Physics), KTH Royal Institute of Technology, Stockholm University, and Uppsala University, Stockholm, Sweden; Department of Mathematics, Stockholm University, Stockholm, Sweden; Santa Fe Institute, Santa Fe, NM 87501 {USA}.
(\email{mason.porter@su.se})
}
}

\usepackage{amsopn}
\usepackage{cite}

\ifpdf
\hypersetup{
  pdftitle={An Approximate-Master-Equation Formulation of the Watts Threshold Model on Hypergraphs},
  pdfauthor={L. A. Keating, K.-I. Goh, and M. A. Porter}
}
\fi

\definecolor{masoncolor}{rgb}{0.98, 0.27, 0.62}
\definecolor{newcolor}{rgb}{0.5, 0.1, 0.98}

\begin{document}

\maketitle

\begin{abstract}
In traditional models of behavioral and opinion dynamics on networks, researchers suppose that interactions occur between pairs of individuals. However, in reality, social interactions also occur in sets of three or more individuals. A common way to incorporate such polyadic interactions is to study dynamical processes on hypergraphs, in which interactions can occur between any number of the individuals. A major current effort is to generalize popular dynamical processes from graphs to hypergraphs. One important model is the Watts threshold model (WTM), which describes a simplistic social spreading process and was extended {recently} by Chen et al.~\cite{chen2025simple} from dyadic networks (i.e., ordinary graphs) to polyadic networks (i.e., hypergraphs). In the present paper, we extend their discrete-time model to continuous time by using approximate master equations (AMEs). By employing AMEs, we are able to very accurately model the mean behavior of the continuous-time system.
We {reduce} the high-dimensional AME system to a system of three coupled {ordinary} differential equations without any detectable loss of accuracy. This much lower-dimensional system is {computationally faster}
to solve numerically, and it is also easier to interpret. We linearize the reduced AME system and derive a cascade condition, which allows us to determine when a large spreading event occurs. We then examine our reduced {AME system} on a contact hypergraph of a French primary school and on a hypergraph of computer-science coauthorships. We find that our full and reduced AME systems give reasonably good descriptions of the mean time-dependent fraction of active nodes in the polyadic WTM dynamics on these empirical networks.
\end{abstract}





\section{Introduction}

In social spreading processes, such as the spread of behavior \cite{centola2010spread} and the adoption of fads \cite{sprague2017evidence}, individuals are sometimes more likely to adopt
 information or other things 
 {when many of their neighbors in a social network} have already adopted it \cite{lehmann2018}. Such social reinforcement occurs in social contagions (which are thus sometimes called ``complex contagions") and distinguishes them from phenomena like the spread of infectious diseases, where it is common to assume that different interactions are independent of each {other~\cite{porter2016,pastor2015}.}
 One popular (but simplistic) model of a social contagion is the Watts threshold model (WTM) {\cite{watts2002simple}.\footnote{See also the threshold models in Kempe et al.~\cite{kkt2003}, prior studies of threshold models in the social sciences \cite{granovetter1978,valente-book}, and prior research on so-called ``bootstrap percolation" in physics~\cite{bootstrap1979}.}} 
In the WTM, each node of a network has a fixed adoption threshold and nodes can be in one of two states. One can interpret these states as {labeling} ``inactive" nodes versus ``active" nodes. Initially, most nodes are inactive, but a small seed fraction of the nodes are active. Subsequently, at each discrete time step, a node becomes active if the fraction of its neighbors that are active is at least as large as its threshold. One can interpret this change of a node from the inactive state to the active state as the peer pressure from its neighbors overcoming its inertia, which is represented by its threshold \cite{melnik2013}. Once a node becomes active, it remains active forever. {The WTM is simplistic, but it} has been generalized in numerous ways \cite{lehmann2018}, including by incorporating antiestablishment (i.e., ``hipster") {individuals} \cite{juul2019}, multiple stages of activation \cite{melnik2013}, random edge weights \cite{hurd2013watts}, time-dependent relationships between nodes \cite{karimi2013threshold}, multiplex relationships \cite{brummitt2012multiplexity}, signed relationships \cite{lee2023signed}, polyadic relationships \cite{chen2025simple}, and many other ideas.

In the present paper, we study an extension of the {dyadic} WTM by Chen et al.~\cite{chen2025simple} to hypergraphs. Hypergraphs allow one to encode polyadic (i.e., ``higher-order") interactions {{between}} individuals. Traditionally, researchers have represented social interactions in networks by connecting nodes in a pairwise (i.e., dyadic) manner. However, although this approach has led to rich variety of insights in a wealth of applications, many interactions in real life are not dyadic~\cite{battiston2020networks,bick2023,battiston2021physics,bianconi2021higher,battiston2025}. For example, it seems more appropriate to represent a conversation between a group of friends as a {hyperedge, which} is attached simultaneously to all individuals in the group, {than} as a set of pairwise edges. In a hypergraph, each entity of a network is a node and nodes are adjacent via hyperedges, which can consist of any positive number of nodes. {Hyperedges thereby encode group interactions. In \cref{fig:hyper_schematic}, we show an illustration of a small hypergraph.} In Chen et al.'s hypergraph extension of the WTM~\cite{chen2025simple}, each node has a threshold and each group (i.e., each hyperedge) also has a threshold. An inactive node activates if the fraction of its groups that are active is at least its threshold. Similarly, a hyperedge activates if the fraction of its constituent nodes that are active is at least its threshold. Chen et al. considered these node and hyperedge activations in discrete time. In the present paper, we study their double-threshold hypergraph WTM in continuous time using approximate master equations (AMEs). 
{In a prior study, Gleeson~\cite{gleeson2013} used AMEs to study a continuous-time extension of the dyadic WTM.} 
{{In the present paper, we use AMEs to analyze a continuous-time WTM that incorporates polyadic interactions between individuals.}}

It has become vogue to model spreading processes on networks with polyadic {interactions~\cite{ferraz2024contagion, majhi2022dynamics,battiston2025,carletti2020dynamical}.} 
One can represent polyadic interactions in a variety of {ways~\cite{battiston2020networks,peixoto2026}}. For example, in a simplicial complex, {whenever a hyperedge is attached to a set of nodes, there is necessarily also a} hyperedge that is attached to any subset of {those} nodes. This ``downward-closure" requirement of simplicial complexes allows one to interpret them as special types of hypergraphs, which one can also use to represent polyadic networks~\cite{battiston2020networks,bick2023,battiston2021physics,bianconi2021higher}. The choice to use a simplicial-complex description instead of a more general hypergraph representation is usually for mathematical convenience, as it leads to some beautiful and useful mathematical theory~\cite{bick2023}. However, it forces restrictions on network structure that do not seem to appropriately model most real-world networks.
Therefore, we use hypergraphs in the present paper.
The choice between employing hypergraphs or simplicial complexes also has important ramifications for the qualitative behavior of dynamical processes on polyadic networks.
For example, Zhang et al.~\cite{zhang2023higher} showed for a Kuramoto coupled-oscillator model on polyadic networks that stronger coupling in the higher-order edges (which connect three or more nodes) leads to more stable synchronization in random simplicial complexes but to less stable synchronization in random hypergraphs.

To give further context for our study, we highlight several investigations of spreading processes and opinion dynamics on polyadic networks~\cite{caldarelli2025, starnini2025,battiston2025}. {Iacopini} et al.~\cite{iacopini2019simplicial} examined a mean-field model of susceptible--infected--susceptible (SIS) contagions on simplicial complexes{,} and Landry and Restrepo~\cite{landry2020effect} examined a mean-field model of SIS dynamics on networks with dyadic and triadic interactions. St-Onge et al.~\cite{st2021master,stonge2022} studied SIS dynamics on hypergraphs using AMEs to explore {mesoscale} localization and seeding strategies for social contagions, and Burgio et al.~\cite{burgio2023adaptive} examined adaptive hypergraph dynamics using AMEs. {Additionally, Kim et al. studied the impact of hyperedge nestedness on both SIS contagions \cite{kim2023contagion} and simplicial contagions \cite{kim2026} on hypergraphs.}
 {{Bret\'{o}n-Fuertes} et al.~\cite{breton2025}} {modeled the emergence of systematic corruption by studying influence dynamics on hypergraphs, finding that polyadic interactions induce sharp transitions between fully honest and fully corrupt regimes}. Opinion models that have been studied on polyadic networks include voter models~\cite{kuehn2020,kim2024competition}, bounded-confidence models~\cite{hickok2022,schawe2022}, models for consensus dynamics \cite{neuhauser2020multibody}, and a model that tracks the states of both nodes and groups~\cite{sampson2024}.


{In the present paper,} we use AMEs to study a continuous-time variant of Chen et al.'s double-threshold hypergraph WTM~\cite{chen2025simple}. In an AME, one examines a dynamical process on a network by tracking the evolution of the fraction of nodes or edges (or hyperedges, in polyadic situations) in each network state. For dyadic networks, AMEs usually track the exact dynamics of a focal node, but we use a mean-field approximation of the dynamics of its neighbors \cite{gleeson2011high,gleeson2013}. {In a study of} the dyadic WTM, Gleeson \cite{gleeson2013} used an AME to accurately capture the fraction of active nodes and the fraction of discordant edges (i.e., edges between active and inactive nodes) as {functions} of time. By contrast, the mean-field and pair approximations that he employed did not accurately capture {their dynamics.}
The double-threshold polyadic WTM that we study has thresholds for both nodes and hyperedges (i.e., groups). Therefore, we expect that it is
insufficient to use a mean-field approximation for the node dynamics, as has been employed in prior {research} on other contagion models on hypergraphs~\cite{st2021master,stonge2022}. We use AMEs for both node dynamics and group dynamics, as was also done recently by Burgio et al.~\cite{burgio2023adaptive} in a study of contagion dynamics on adaptive hypergraphs.

{Gleeson \cite{gleeson2013} showed for the dyadic WTM that} employing an appropriate ansatz allows one to reduce the full AME system to two coupled ordinary differential equations (ODEs). In the present paper, we generalize this reduction to the double-threshold polyadic WTM and use a pair of ansatzes to reduce the AME system to three coupled ODEs without any detectable loss {of} accuracy. Leveraging this much lower dimensionality, we {analyze} the linear stability of the reduced AME system and efficiently solve it numerically.

Our paper proceeds as follows. In \Cref{sec:model}, we describe the double-threshold polyadic WTM and set up the corresponding system of AMEs. In \Cref{sec:reduced_dim}, we reduce the AME system to a set of three coupled ODEs using a pair of ansatzes. In \Cref{sec:cascade_condition}, we derive a cascade condition. In \Cref{sec:empirical}, we show that the reduced AME system for the polyadic WTM gives a reasonably good description of the mean time-dependent fraction of active nodes on empirical networks. Finally, in \Cref{sec:discussion}, we conclude and discuss our results. In \Cref{sec:chen}, we compare {the steady-state fraction of active nodes in the reduced AME system as a function of the mean number of groups per node (i.e., the {degree}\footnote{
{In the present paper, we use the term ``degree" for the number of groups that a node is in. In hypergraphs, it is also common to use the term ``hyperdegree" for this quantity.}})
and as a function of the mean group size (i.e., {the} mean hyperedge size)} with those of Chen et al.~\cite{chen2025simple}. Our code is {available} at \url{https://github.com/leahkeating/threshold_AME}.


\section{A Polyadic Threshold Model with Both Node Thresholds and Group Thresholds}  \label{sec:model}

We study a continuous-time extension of the discrete-time double-threshold polyadic threshold model of Chen et al.~\cite{chen2025simple}. In \Cref{sec:chen}, we illustrate that there is excellent agreement between 
the steady-state fraction of active nodes in our model and in Chen et al.'s model. {These polyadic threshold models are extensions of the dyadic}
WTM~\cite{watts2002simple} to hypergraphs. 

\begin{figure}[h]
\medskip
    \centering
    \includegraphics[width=0.7\linewidth]{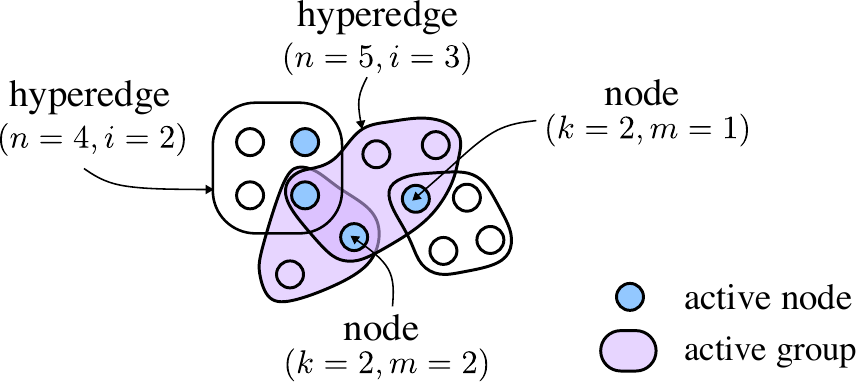}
    \caption{A small hypergraph with active nodes shaded in blue and active hyperedges shaded in purple. We show the state of two hyperedges, where $n$ is the number of nodes in the hyperedge and $i$ is the number of active nodes in the hyperedge. We show hyperedges with $(n = 5, i = 3)$ and $(n = 4, i = 2)$. We also show the states of two nodes, where $k$ is the degree of the node and $m$ is the number of active groups (i.e., {{hyperedges}}) to which a node is attached. We show nodes with $(k = 2, m = 2)$ and $(k = 2, m = 1)$.
    }
    \label{fig:hyper_schematic}
\end{figure}

In a hypergraph, nodes are adjacent to each other via hyperedges (i.e., groups) that can consist of {a} nonnegative number of nodes. The ``size" of a hyperedge{, which one can interpret as a group of nodes,} is the number of nodes that are attached to it. {The ``degree" $k_{\ell}$ of a node $x_{\ell}$ is its number of attached hyperedges. In \cref{fig:hyper_schematic}, we show a small hypergraph.} 

In our double-threshold WTM, we assign ``classes" to both the nodes and the hyperedges. Two nodes (respectively, hyperedges) are in the same class if they have the same degree (respectively, size) and {same} threshold. Each node $x_{\ell}$ is in a class $\vec{k}_{\ell} = \{k_{\ell}, \sigma_{k_{\ell}}\}$ that is described by {its degree $k_{\ell}$ and its threshold $\sigma_{k_{\ell}}$.} Each hyperedge $y_j$ is in a class $\vec{n}_j = \{n_j, {\tau_{n_j}}\}$ that is described by its size $n_j$ and {its} threshold 
{$\tau_{n_j}$}. {Let $\{g_k\}$ denote the degree distribution of the nodes of a network, $\langle k \rangle$ denote the network's mean degree, $\{p_n\}$ denote the group-size distribution of the hyperedges of a network, and $\langle n \rangle$ denote the network's mean group size (i.e., mean hyperedge size).} With the above notions of classes, we are assuming that all nodes with the same degree $k$ are in the same class $\vec{k}$ and that all hyperedges of the same size $n$ are in the same class $\vec{n}$. 
{We are assuming that} all degree-$k$ nodes have the same threshold $\sigma_k$ and {that} all size-$n$ hyperedges have the same threshold {$\tau_n$}.

We initialize the dynamics by uniformly randomly selecting a fraction $\rho_0$ of nodes to be active. We then set groups (i.e., hyperedges) to be active if the {fraction} {$\nicefrac{i}{n}$} of active nodes in the group is at least the group's threshold {$\tau_n$} {(i.e., if the number of active nodes is ${i} \geq {n}\tau_n$).} 
The system evolves asynchronously, although multiple nodes update their states simultaneously, with a selected node activating if its fraction of active groups is at least $\sigma_k$ and a selected group activating if the fraction of active nodes in the group is at least {$\tau_n$}. In this asynchronous scheme, each node and {each} hyperedge has a chance to update its state {approximately} {once on average in each time} {unit.} {Specifically}, in each time interval of duration $\Delta t$, a fraction $\Delta t$ of the nodes and {a fraction $\Delta t$} of the hyperedges {can}
update their states. Nodes and groups update their states based on the states of other nodes and groups at the end of the previous time step.

\begin{figure}[h]
    \centering
    \subfloat[]{\includegraphics[width = 0.5\textwidth]{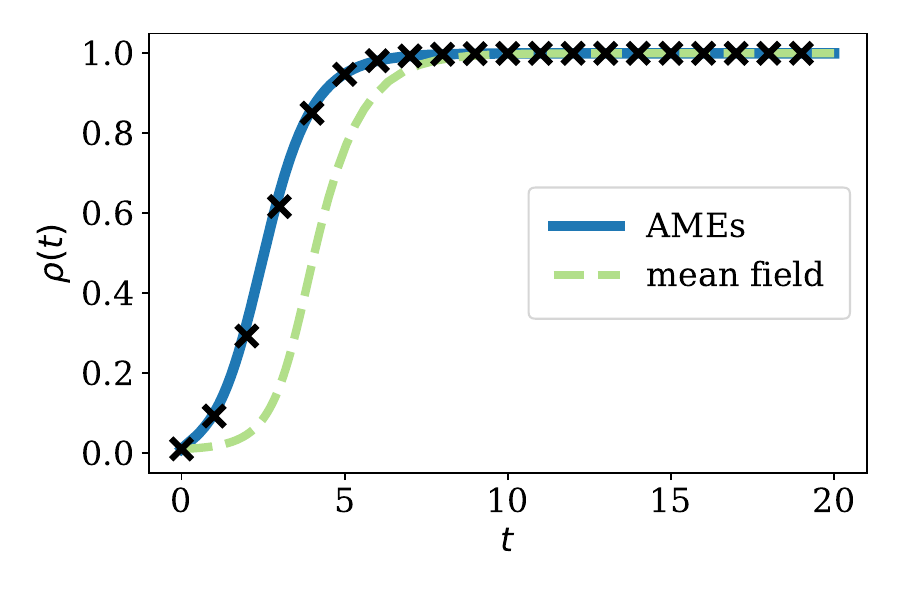}}
    \subfloat[]{\includegraphics[width = 0.5\textwidth]{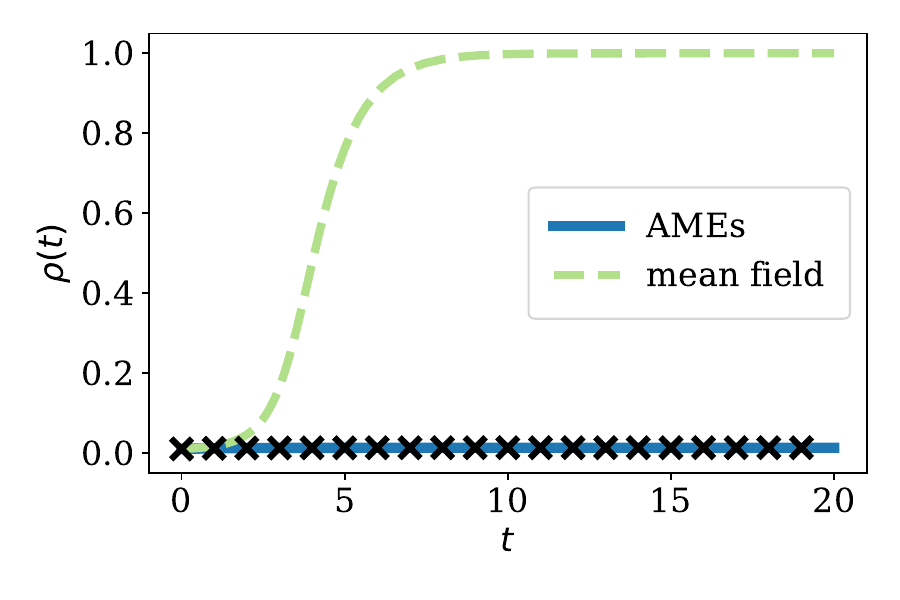}}
    \caption{Comparison of numerical solutions of our mean-field approximation {(\ref{eq:mf_rho_k}, \ref{eq:mf_zeta_n})}
    (dashed green curves) and the full AME system {\cref{eq:fni}--\cref{eq:alpha}}
    (solid blue curves) with the mean active node fraction $\rho(t)$ (black markers) of our continuous-time polyadic double-threshold WTM {for} two sets of 500 WTM simulations on 10,000-node configuration-model hypergraphs. 
    {(We describe how we generate these hypergraphs in \Cref{sec:model}.)}
     We generate a new network for each simulation. In both sets of simulations, the group-size distribution of the hyperedges is $p_n = \delta_{3,n}$, the degree distribution of the nodes is $g_k = \delta_{4,k}$,
     and the fraction of initially {active} nodes is $\rho_0 = 0.01$.
           In (a), the node threshold is $\sigma_k = 0.2$ for all $k$ and the group threshold is ${\tau_n} = 0.3$ for all $n$. In (b), the node threshold is $\sigma_k = 0.3$ for all $k$ and the group threshold is ${\tau_n} = 0.3$ for all $n$.
     }
    \label{fig:mf_vs_ames}
\end{figure}

We derive a mean-field approximation of the polyadic double-threshold WTM dynamics by generalizing {Eq.~(5)} of Ref.~\cite{gleeson2013} to hypergraphs. We obtain the mean-field equations 
\begin{align}
    \frac{d}{dt}\rho_k &= (1 - \rho_k)\sum_{m = 0}^k \gamma(k,m) B_{k,m}(\xi) \,,
    \label{eq:mf_rho_k} \\
    \frac{d}{dt}\zeta_n &= (1 - \zeta_n)\sum_{i = 0}^n \beta(n,i)B_{n,i}(\omega) \,,
    \label{eq:mf_zeta_n}
\end{align}
where $\rho_k(t)$ is the fraction of degree-$k$ nodes that are active at time $t$, the quantity $\zeta_n(t)$ is the fraction of size-$n$ groups that are active at time $t$, the function $\gamma(k,m)$ is the activation rate of the nodes (see \cref{eq:gamma} below), 
{$\beta(n,i)$} is the activation rate of the groups (see \cref{eq:beta} below), 
{{$\xi(t) = \frac{1}{\sum_n n p_n}\sum_n np_n\zeta_n(t)$} is the probability that a uniformly randomly selected group of a node is active, $\omega(t) = \frac{1}{\sum_k k g_k}\sum_k kg_k\rho_k(t)$} is the probability that a uniformly randomly selected node {in}
a group is active, and $B_{a,b}(x)$ is the binomial probability $\binom{a}{b}x^b (1 - x)^{a - b}$. We calculate the fraction $\rho(t)$ of active nodes at time $t$ using the expression $\rho(t) = \sum_k g_k \rho_k (t)$. For the mean-field model {(\ref{eq:mf_rho_k}, \ref{eq:mf_zeta_n}),}
the initial conditions are $\rho_k (0) = \rho_0$ and $\zeta_n(0) = \sum_{i\geq n{\tau_n}}B_{n,i}(\rho_0)$, where $\rho_0$ is the fraction of initially active nodes. In \cref{fig:mf_vs_ames}, we show numerical solutions of the {mean-field approximation} {(\ref{eq:mf_rho_k}, \ref{eq:mf_zeta_n})}.
{The} mean-field dynamics are very different from the dynamics of the polyadic double-threshold WTM.


To model the polyadic WTM dynamics more accurately than we can with the mean-field approximation {(\ref{eq:mf_rho_k}, \ref{eq:mf_zeta_n})},
we use approximate master equations (AMEs) to track the densities of nodes and hyperedges in specific states. As in the mean-field approximation, the AMEs rely on the underlying assumption that there are no correlations between the sizes of groups and the degrees of their {{constituent}} nodes. 
{We} track the fraction $s_{k,m}(t)$ of degree-$k$ nodes that are inactive at time $t$ and in $m$ active groups and the fraction $f_{n,i}(t)$ of size-$n$ groups that are inactive at time $t$ and have $i$ active nodes. The fraction of degree-$k$ nodes that are inactive at time $t$ is $\sum_{m = 0}^{k} s_{k,m}(t)$, and the fraction of size-$n$ groups that are inactive at time $t$ is $\sum_{i = 0}^n f_{n,i}(t)$. We track the fraction $\rho(t)$ of active nodes at time $t$ by calculating $1 - \sum_{k = 0}^\infty g_k \sum_{m = 0}^k s_{k,m}(t)$, where the degree distribution $\{g_k\}$ gives the probability that a uniformly randomly selected node has degree $k$ for each value of $k$.
Similarly, the fraction of active groups at time $t$ is $1 - \sum_{n = 0}^\infty p_n \sum_{i = 0}^n f_{n,i}(t)$, where the hyperedge-size distribution $\{p_n\}$ gives the probability that a uniformly randomly selected hyperedge (i.e., group) has size $n$.

The probability that a size-$n$ group has $i$ active nodes at time $t = 0$ is 
$B_{n,i}(\rho_0) = \binom{n}{i}\rho_0^i (1 - \rho_0)^{n - i}$. Such a group is inactive if $i$ is below the node's threshold. 
The initial fraction of size-$n$ groups that are inactive and have $i$ active nodes is 
\begin{equation}
    f_{n,i}(0) = \begin{cases}
   			 B_{n,i}(\rho_0) & \text{if } i < n {\tau_n}  \\
   			 0 & \text{otherwise} \,.
   		 \end{cases}
    \label{eq:fni_0}
\end{equation}
To calculate the initial fraction $s_{k,m}(0)$ of degree-$k$ nodes that are in $m$ active groups, we use the facts that {(i)} a uniformly randomly selected node is inactive with probability $1 - \rho_0$ and {(ii)} the probability {that} an inactive degree-$k$ node is in exactly $m$ active groups is $B_{k,m}(\phi_0)$, where $\phi_0$ is the probability that a uniformly randomly selected group of an inactive node is initially active. We find that
\begin{equation}
    s_{k,m}(0) = (1 - \rho_{0})B_{k,m}\left(\phi_0\right) 
\end{equation}
and 
\begin{equation}
    \phi_0 = \frac{\sum_{n = 0}^\infty n p_n \sum_{i = 0}^{n - 1} B_{n - 1, i}(\rho_0)\mathbbm{1}_{\left[i\geq n {\tau_n}\right]}}{\sum_{n = 0}^\infty n p_n} \,,
    \label{eq:phi_0}
\end{equation}
where the indicator function $\mathbbm{1}_W$ takes the value $1$ on the set $W$ and the value $0$ everywhere else.

We now detail the full system of AMEs that describe the WTM dynamics in terms of $s_{k,m}(t)$ and $f_{n,i}(t)$. The rate of change of the fraction $f_{n,i}(t)$ of size-$n$ groups that are inactive and have $i$ active nodes at time $t$ is {given by the master equation} 
\begin{equation}
    \frac{df_{n,i}}{dt} = -\beta(n,i)f_{n,i} + (n - i + 1)\eta f_{n,i - 1} - (n - i)\eta f_{n,i} \,.
    \label{eq:fni}
\end{equation}

The first term ($-\beta(n,i)f_{n,i}$) on the right-hand side of \cref{eq:fni} accounts for a size-$n$ group activating when its threshold is met or exceeded. The group activation function $\beta(n,i)$ in this term is
\begin{equation}
    \beta(n,i) = \begin{cases}
        1 & \text{if }{i\geq n}{\tau_n} \\
        0 & \text{otherwise} \,.
    		\end{cases}
    \label{eq:beta}
\end{equation} 
The second term ($+(n - i + 1)\eta f_{n,i - 1}$) on the right-hand side of \cref{eq:fni} accounts for the activation of one of the nodes in a size-$n$ group with $i - 1$ active nodes to yield a size-$n$ group with $i$ active nodes. The last term ($-(n - i)\eta f_{n,i}$) on the right-hand side of \cref{eq:fni} accounts for the activation of a node in a size-$n$ group with $i$ active nodes to yield a size-$n$ group with $i + 1$ active nodes. {The variable 
\begin{equation} \label{eq:eta}
    \eta(t) = \frac{\sum_{k = 0}^{{\infty}}g_{k}\sum_{m = 0}^{k}(k - m)\gamma(k,m)s_{k,m}(t)}{\sum_{k = 0}^\infty g_{k}\sum_{m = 0}^{k}(k - m)s_{k,m}(t)}
\end{equation} 
is the mean-field activation rate of an inactive node in an inactive group at time $t$.}
The quantity $\gamma(k,m)$ (see \cref{eq:gamma} below) in \cref{eq:eta} is the activation function for the nodes.

The AMEs that track the fraction $s_{k,m}$ of degree-$k$ nodes that are in $m$ active groups and are themselves inactive are 
\begin{equation}
    \frac{ds_{k,m}}{dt} = -\gamma(k,m)s_{k,m} + (k - m + 1)\alpha s_{k,m - 1} - (k - m)\alpha s_{k,m} \,.
    \label{eq:skm}
\end{equation}
The first term ($-\gamma(k,m)s_{k,m}$) on the right-hand side of \cref{eq:skm} accounts for a degree-$k$ node in $m$ active groups activating when its threshold is met or exceeded. The prefactor function $\gamma(k,m)$ in this term is
\begin{equation}
    \gamma(k,m) = \begin{cases}
        1& \text{if }{m\geq k}\sigma_{k} \\
        0&\text{otherwise} \,.
    \end{cases}\label{eq:gamma}
\end{equation}
The second term ($+(k - m + 1)\alpha s_{k,m - 1}$) on the right-hand side of \cref{eq:skm} accounts for the group activation of an inactive degree-$k$ node in $m - 1$ active groups to yield an inactive degree-$k$ node in $m$ active groups. This transition causes an increase in $s_{k,m}$. 
{The variable 
\begin{equation} \label{eq:alpha}
    \alpha(t) = \frac{\sum_{n = 0}^{{\infty}}p_{n}\sum_{i = 0}^{n}(n - i)\beta(n,i)f_{n,i}(t)}{\sum_{n = 0}^\infty p_{n}\sum_{i = 0}^{n}(n - i)f_{n,i}(t)}
\end{equation}
is the {mean-field} activation rate of an inactive group of an inactive node at time $t$.}
The last term ($-(k - m)\alpha s_{k,m}$) on the right-hand side of \cref{eq:skm} accounts for the group activation of an inactive degree-$k$ node in $m$ active groups to yield an inactive degree-$k$ node in $m + 1$ active groups. This transition causes a decrease in $s_{k,m}$.

\begin{figure}[h]
    \centering
    \includegraphics[width=\linewidth]{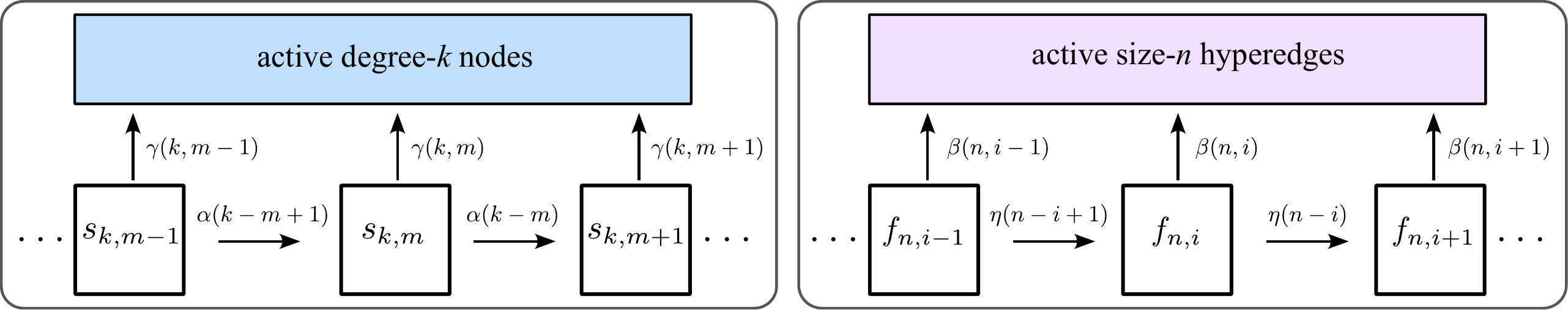}
    \caption{
    A visual representation of the transitions into and out of {node states and hyperedge (i.e., group) states}
    that we track using the AMEs {(\ref{eq:fni})--(\ref{eq:alpha})}. 
    In (a), we show all possible transitions to and from the inactive node class $(k,m)$, where $s_{k,m}$ is the fraction of degree-$k$ nodes that are inactive and in $m$ active groups. A uniformly randomly selected neighbor of an inactive node activates at rate $\alpha$, and inactive nodes activate at rate $\gamma(k,m)$, which equals $1$ if a node's threshold is met and equals $0$ if it is not met. In (b), we show all possible transitions to and from the inactive {group}
    class $(n,i)$, where $f_{n,i}$ is the probability that a size-$n$ {group}
    is inactive and has $i$ active nodes. A uniformly randomly selected inactive node {in an inactive group}
    activates at rate $\eta$, and {an inactive $n$-node group with $i$ active nodes}
    activates at rate $\beta(n,i)$, which equals $1$ if a {group's}
    threshold is met and equals $0$ if it is not met.
    }
    \label{fig:node_edge_state_transitions}
\end{figure}

\begin{figure}[h]
    \centering
    \subfloat[]{\includegraphics[width = 0.5\textwidth]{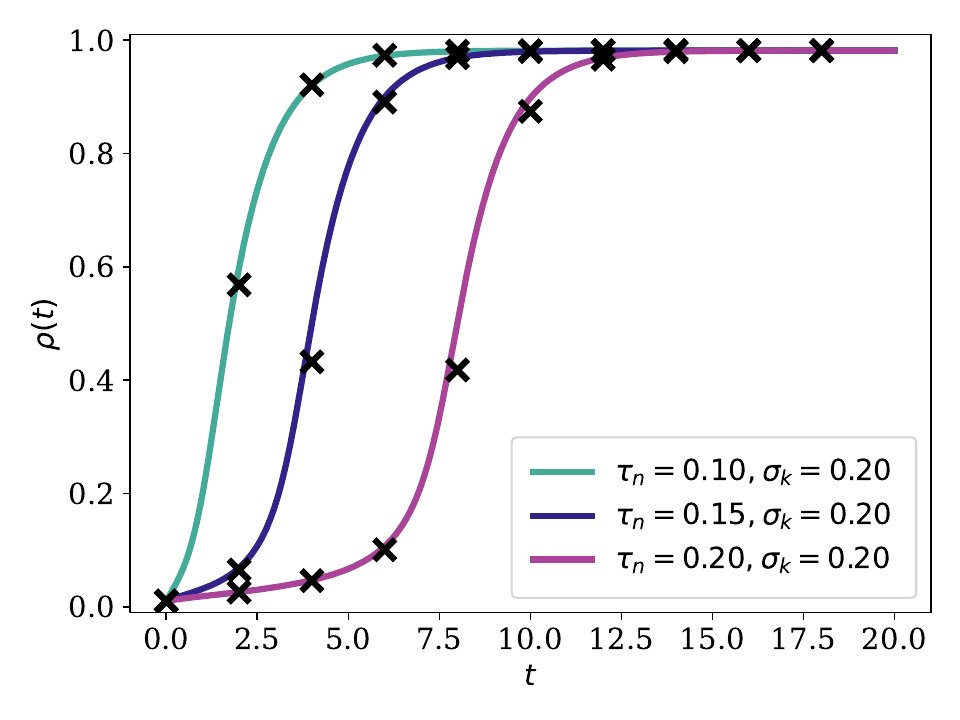}}
    \subfloat[]{\includegraphics[width = 0.5\textwidth]{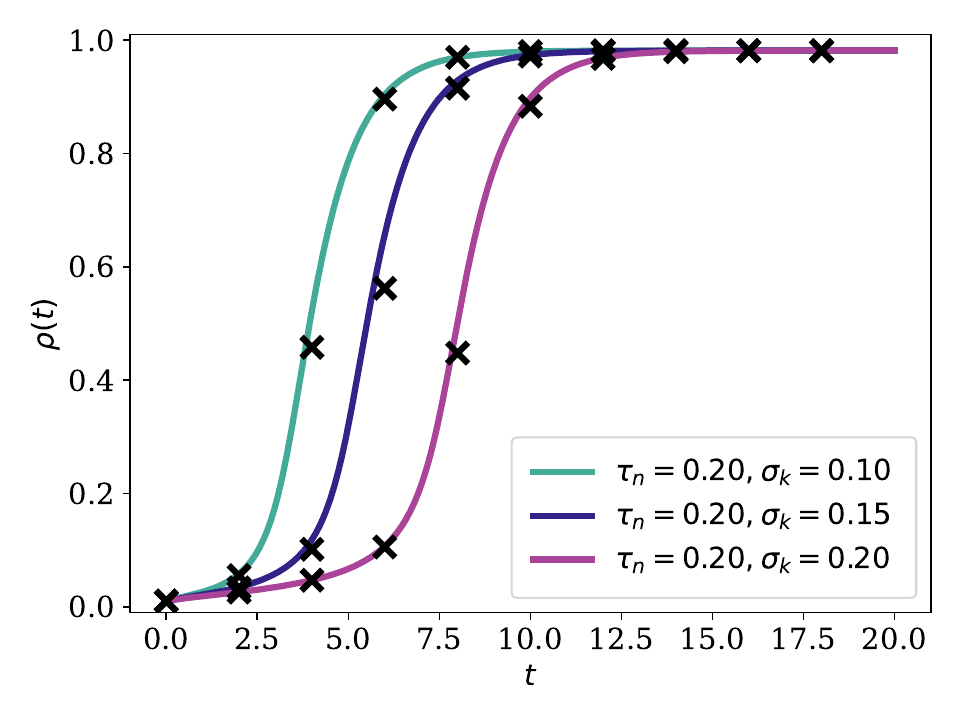}}
    \caption{The dependence of the fraction $\rho(t)$ of active nodes at time $t$ on (a) the group threshold ${\tau_n}$ and (b) the node threshold $\sigma_k$.
    Each curve is a numerical solution of the full AME system \cref{eq:fni}--\cref{eq:alpha},
    where $\rho(t) = 1 - \sum_k g_k \sum_m s_{k,m}(t)$. The black markers are means of 500 simulations on 500 different synthetically-generated hypergraphs with 50,000 nodes. For each curve, the initial fraction of active nodes is $\rho_0 = 10^{-2}$. The group-size distribution is $p_n \sim \mathrm{Pois}(8)$ {(where the notation $\sim$ indicates that $p_n$ follows the indicated probability distribution)}, and the degree distribution of the nodes is $g_k \sim \mathrm{Pois}(4)$. In (a), we fix the node {threshold at} 
    $\sigma_k = 0.2$ and vary the group threshold ${\tau_n}$. As we decrease $\sigma_k$, we observe that $\rho(t)$ increases more rapidly. In (b), we fix the group {threshold at}
    $\tau_n = 0.2$ and vary the node threshold $\sigma_k$. As we decrease $\sigma_k$, we observe that $\rho(t)$ increases more rapidly. By comparing panels (a) and (b), we observe that the impact of decreasing group thresholds differs from the impact of decreasing node thresholds.
    }
    \label{fig:full_soln}
\end{figure}

In \cref{fig:node_edge_state_transitions}, we illustrate each of the possible transitions to and from the inactive node state $s_{k,m}$ and to and from the inactive {group}
state $f_{n,i}$. In \cref{fig:full_soln}, we show the solution $\rho(t)$ of the full AME system \cref{eq:fni}--\cref{eq:alpha} on hypergraphs in which both {the node degrees and the group sizes}
follow independent Poisson distributions for {three fixed node thresholds} $\sigma_k$ and {three fixed group thresholds} ${\tau_n}$. 
{In a Poisson distribution $\mathrm{Pois}(\lambda)$ with parameter $\lambda$, the probability that a random variable $X$ has the value $x$ is
\begin{equation}
    \frac{\lambda^x e^{-\lambda}}{x!} \,.
\end{equation}
In our investigation, we always assume that each node has the same threshold and that each group has the same threshold, although our model allows one to consider heterogeneous thresholds.}
{In \cref{fig:full_soln}(a), we fix the node threshold at $\sigma_k = 0.2$ and vary the group {threshold. In} \cref{fig:full_soln}(b), we fix the group threshold at $\sigma_n = 0.2$ and vary the node threshold.} 


In each of our simulations, we generate a hypergraph with a prescribed number of nodes, degree distribution, and {group-size (i.e., hyperedge-size)}
distribution from a configuration model.
 We then run the WTM dynamics with asynchronous updates of the node states and {group}
states. We consider 500 realizations for each of our numerical experiments. The configuration model{\footnote{{Chodrow~\cite{chodrow2020configuration} studied a similar hypergraph configuration model.}} that we use} to generate hypergraphs is an extension of the configuration-model graphs in \cite{hebert2010propagation} to hypergraphs. (As discussed by Fosdick et al.~\cite{fosdick2018} in the context of ordinary graphs, different specifications of configuration-model networks can lead to meaningful differences in the properties of such networks.) To construct a hypergraph, we generate a list of $N$ node labels, where $N$ is the number of nodes in the network. We sample the degree of each node from the distribution $\{g_k\}$ and store these values in a separate list. We then introduce hyperedges {one by one} by sampling each hyperedge size from the distribution $\{p_n\}$ until the sum of the hyperedge sizes equals the sum of the node degrees 
(i.e., $\sum_{\ell} n_{\ell} = \sum_{\ell} k_{\ell}$). We do not fix the total number of hyperedges. If the sum of the {hyperedge sizes ever exceeds the sum of the node degrees}, we restart the process of adding hyperedges {one by one}. Once we have our set of hyperedges, we {(i)} create a list of node identities with $k_{\ell}$ copies of node ${\ell}$, {(ii)} create a list of hyperedge identities with $n_j$ copies of hyperedge $j$, and {(iii)} shuffle both lists to obtain uniformly random orders in each list. 
By construction, both of these lists have the same length. We assign nodes to hyperedges by matching 
{nodes} to corresponding hyperedges in each list. 
It is possible for a node to occur more than once in the same hyperedge, but our networks are large (they usually have more than 10,000 nodes), so this situation is rare.
(In our calculations, we ignore any repeated nodes.) This process yields a hyperedge list, which encodes which nodes are in each hyperedge. From this list, we generate a hypergraph using the XGI {\sc Python} software package~\cite{Landry_XGI_2023}. 


{To} simulate dynamics on a hypergraph, we divide each unit of time into ${1}/{\Delta t}$ smaller time {steps. In each} time step, we update a fraction $\Delta t$ of the nodes uniformly at random and we then update a fraction $\Delta t$ of the hyperedges. In our examples, we use $\Delta t = 0.1$. In one time step, it is possible that some nodes are never selected and also that some nodes are selected more than once. However, we expect {these scenarios to have negligible impact on our results.}
 {When}
 $\Delta t = 0.1$, the probability that {we select a specific node exactly once in a specified time unit is {$(\nicefrac{1}{\Delta t})(1 - \Delta t)^{\nicefrac{1}{\Delta t} - 1}\Delta t \approx 0.39$}{. If} we} decrease $\Delta t$, this probability decreases. {Additionally, for $\Delta t = 0.1$, the probability that we do not select a specified node to update its state in a time unit} is {$(1 - \Delta t)^{\nicefrac{1}{\Delta t}} \approx 0.34$}{. This} probability decreases as we decrease $\Delta t$. We obtain excellent agreement between WTM simulations and the full AME system {\cref{eq:fni}--\cref{eq:alpha}} (see \cref{fig:full_soln}), but they are not exactly the same. 
 This discrepancy may {arise from our use of} \textit{approximate} master equations{. In our approximation, we} track the activation of a group that includes a node (through the function $s_{k,m}(t)$) and the activation of a node that belongs to a group (through the function $f_{n,i}(t)$) using {the mean-field quantities} $\alpha(t)$ and $\eta(t)$.


\section{Reduced AME Equations} \label{sec:reduced_dim}

The full AME system \cref{eq:fni}--\cref{eq:alpha} is typically very high-dimensional when the maximum group size and/or the maximum degree are large. It is thus slow to numerically solve this system and challenging to analytically solve it. This situation motivates us to reduce the full AME system to 
{a} more numerically and analytically tractable system. For example, we do not know how to derive a cascade condition from the full AME system, but we are able to derive an approximate cascade condition (see \Cref{sec:cascade_condition}) from a reduced AME system (see \cref{eq:rho_dot}--\cref{eq:phi_dot} below).

\begin{figure}[h!]
    \centering
    \subfloat[]{\includegraphics[width=0.45\textwidth]{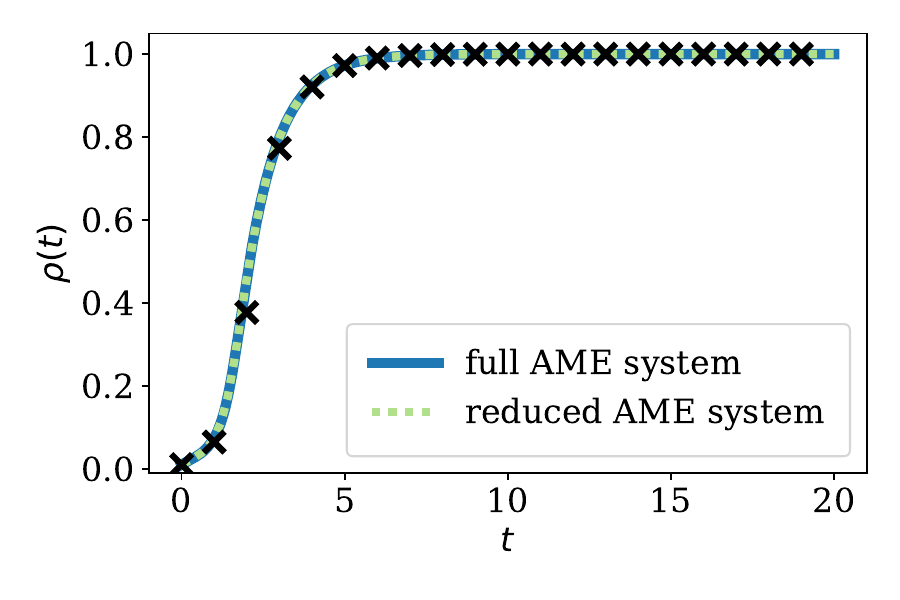}}
    \subfloat[]{\includegraphics[width=0.45\textwidth]{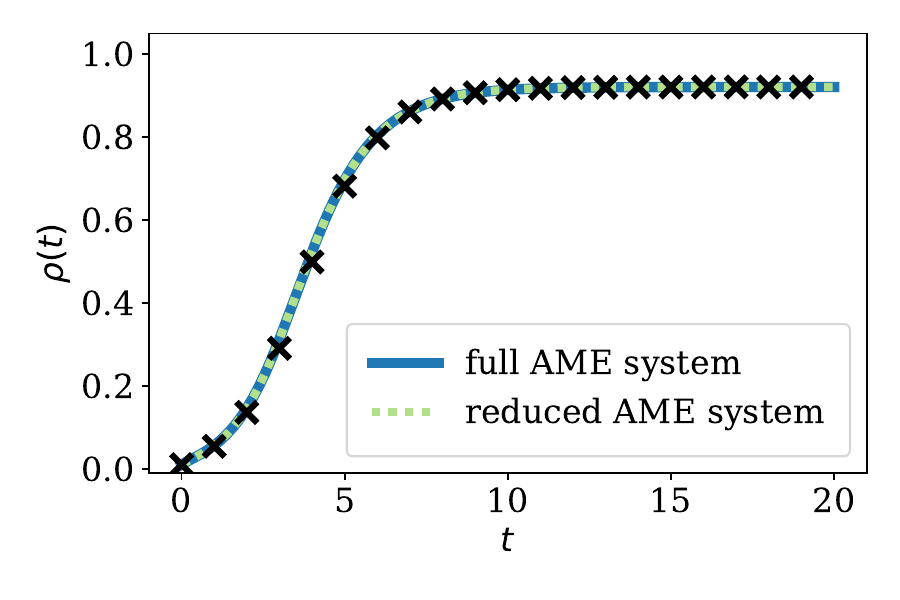}}
    
    \subfloat[]{\includegraphics[width=0.45\textwidth]{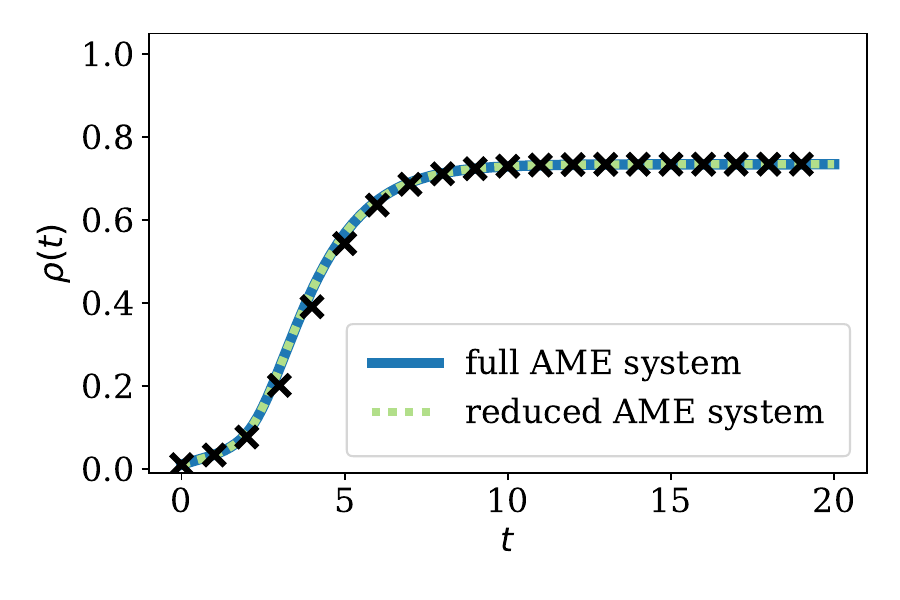}}
    \subfloat[]{\includegraphics[width=0.45\textwidth]{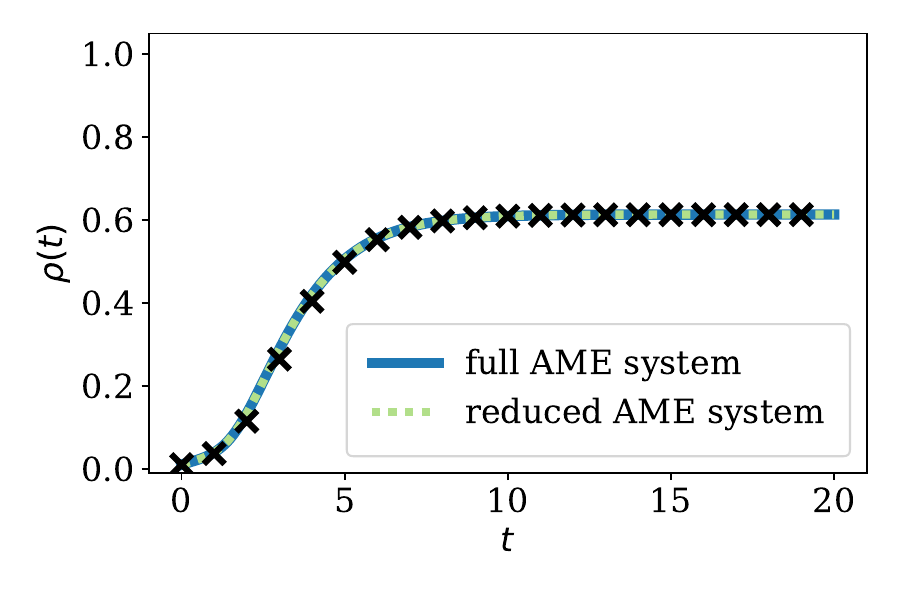}}
    \caption{
    The reduced AME system \cref{eq:rho_dot}--\cref{eq:phi_dot} accurately recovers the solutions of the full AME system \cref{eq:fni}--\cref{eq:alpha}
    for several choices of degree distributions, {group-size}
    distributions, node thresholds, and group thresholds in configuration-model hypergraphs.
    The black markers give the means of 500 simulations on 500 different 10,000-node networks, the solid blue curves give results for the full AME system, and the dotted green curves give results for the reduced AME system.
    (a) An example with degree distribution $g_k\sim\mathrm{Pois}(11)$, hyperedge-size distribution $p_n \sim \mathrm{Pois}(6)$, node threshold $\sigma_k = 0.2$, group threshold ${\tau_n} = 0.1$, and initially active node fraction $\rho_0 = 0.01$.  (b) An example with degree distribution $g_k \sim \mathrm{Pois}(3)$, {group-size distribution} $p_n \sim \mathrm{Pois}(2)$, node threshold $\sigma_k  = 0.2$, group threshold ${\tau_n} = 0.1$, and initially active node fraction $\rho_0 = 0.01$. (c) An example with degree distribution 
{$g_k = \nicefrac{k^{-2.2}}{\sum_{j = 1}^{100}j^{-2.2}}$, {group-size distribution} $p_n = \nicefrac{n^{-2.2}}{\sum_{m = 1}^{100}m^{-2.2}}$,} node threshold $\sigma_k = 0.1$, group threshold ${\tau_n} = 0.1$, and initially active node fraction $\rho_0 = 0.01$. (d) An example 
    with degree distribution {$g_k = \nicefrac{k^{-2.2}}{\sum_{j = 1}^{100}j^{-2.2}}$, {group-size distribution} $p_n = \nicefrac{n^{-2.5}}{\sum_{m = 1}^{100}m^{-2.5}}$,} node threshold $\sigma_k = 0.05$, group threshold ${\tau_n} = 0.1$, and initially active node fraction $\rho_0 = 0.01$. 
    For the heavy-tailed distributions in (c) and (d), {we impose a maximum degree of $100$ and maximum group size of $100$.}
    }
 \label{fig:red_soln}
\end{figure}

Using two ansatzes, which we state below (see {\cref{eq:skm_ansatz,eq:fni_ansatz}}), we reduce the full AME system to a system of three coupled ODEs:
\begin{align}    
    \dot{\rho}(t) &= 1 - \rho(t) - \left(1 - \rho_{0}\right)\sum_{k=0}^{\infty} g_k \left[\sum_{m < k\sigma_k}B_{k,m}(\phi(t))+\delta_{k,0}\right] \,, \label{eq:rho_dot} \\
    \dot{\theta}(t) &=
  	  \begin{cases}
 		   \frac{c_1(1 - \theta(t))(1-\phi(t))-\left(1 - \rho_{0}\right)\sum_{k = 0}^{{\infty}} g_k \sum_{m < k\sigma_k}(k - m)B_{k,m}(\phi(t))}{c_1(1 - \phi(t))} &\text{if }\phi(t) < 1 \\
 		   0&\text{otherwise}\,,
 		   \end{cases}
    \label{eq:theta_dot}\\
    \dot{\phi}(t) &= \begin{cases}
    				\frac{c_2(1 - \theta(t))(1 - \phi(t)) - \sum_{n = 0}^{{\infty}} p_n \sum_{i < n{\tau_n}}(n - i)B_{n,i}(\theta(t))}{c_2(1 - \theta(t))}& \text{if } \theta(t) < 1 \\
   				 0&\text{otherwise} \,,
			\end{cases} 
\label{eq:phi_dot}
\end{align}
where $\rho(t)$ is the probability that a uniformly randomly selected node is active at time $t$, the quantity $\phi(t)$ is the probability that a uniformly randomly selected group of an inactive node is active at time $t$, and $\theta(t)$ is the probability that a uniformly randomly selected node {in}
an inactive group is active at time $t$. We determine the constants $c_1$ and $c_2$ from the initial conditions (see {\cref{eq:c1} and \cref{eq:c2}} below).
In \cref{eq:rho_dot}--\cref{eq:phi_dot}, the quantities $\dot{\phi}(t)$ and $\dot{\theta}(t)$ are independent of $\rho(t)$ and the quantity $\dot{\rho}(t)$ depends on $\phi(t)$. We include \cref{eq:rho_dot} for $\dot{\rho}(t)$ because we are particularly interested in $\rho(t)$. 
The reduction {from the full AME system \cref{eq:fni}--\cref{eq:alpha}} to \cref{eq:rho_dot}--\cref{eq:phi_dot} is exact \cite{gleeson2013} and assumes that we seed active nodes in a hypergraph uniformly at random.\footnote{We have not examined the accuracy of this reduction for any other initial conditions, and this is worth exploring in future efforts. Additionally, for some {group-size}
distributions and degree distributions (though not the ones that we employ), there can be {{discrepancies}} between the reduced AME system {\cref{eq:rho_dot}--\cref{eq:phi_dot}}
and the full AME system \cref{eq:fni}--\cref{eq:alpha} near transition points. 
Such discrepancies arise when the assumptions in the ansatzes \eqref{eq:skm_ansatz} and \eqref{eq:fni_ansatz} are not accurate.}
In \cref{fig:red_soln}, we show that {(\ref{eq:rho_dot})--(\ref{eq:phi_dot})} can give accurate results for $\rho(t)$.
 In this figure, we compare  $\rho(t)$ for the full AME system {\cref{eq:fni}--\cref{eq:alpha},} the reduced AME system {\cref{eq:rho_dot}--\cref{eq:phi_dot}}, and the mean of Monte Carlo simulations of our polyadic WTM for several choices of the degree distribution $\{g_k\}$, the {group-size}
 distribution $\{p_n\}$, the node threshold $\sigma_k$, and the group threshold ${\tau_n}$. 
  {The reduced AME system is more efficient computationally than the full AME system by a factor
of $\mathcal{O}(\max\{K,M\})$ in both memory and time, where $K$ is the maximum degree (i.e., a cutoff) in our numerical computations
and $M$ is the maximum hyperedge size in our numerical computations.}
 {For example,} in \cref{fig:red_soln}(a), a single integration from {time $t = 0$ to time $t = 20$} on an Apple MacBook Pro with {an} M3 processor took $6.74$ seconds for the full AME system {but only} $0.03$ seconds for the reduced {AME} system using the \texttt{solve\_ivp} integrator {in {\sc SciPy} \cite{2020SciPy-NMeth} with the method set to {``\texttt{LSODA}''}.}

To derive the reduced AME system \cref{eq:rho_dot}--\cref{eq:phi_dot} from the full AME system \cref{eq:fni}--\cref{eq:alpha}, we follow the approach of Gleeson \cite{gleeson2013}, who reduced an AME system for the dyadic WTM to two coupled ODEs. {Our} reduced AME system consists of three coupled ODEs for the parameters $\rho(t)$, $\phi(t)$, and $\theta(t)$, whereas the full AME system gives equations of motion for $f_{n,i}$ and $s_{k,m}$ for all group sizes $n$, active-node numbers $i$, degrees $k$, and active-group numbers $m$.{\footnote{{We give a detailaed derivation of the reduced AME system, but people who are familiar with such derivations can proceed directly to \Cref{sec:cascade_condition}.}}} We employ the ansatzes
\begin{align}
    s_{k,m}(t) &= \left[1 - \rho_k (0)\right]B_{k,m}(\phi(t)) \,\, \,\, \text{for } \,\, m < k\sigma_k \,,  \label{eq:skm_ansatz} \\
    f_{n,i}(t) &= B_{n,i}(\theta(t)) \,\, \,\, \text{for }\,\, i < n{\tau_n} \,, \label{eq:fni_ansatz}
\end{align}
where $\rho_k (0) = 1 - \sum_m s_{k,m}(0)$ and $B_{a,b}(x) = \binom{a}{b}x^{b}(1 - x)^{a - b}$. 

We start with the node dynamics. We differentiate  \cref{eq:skm_ansatz} with respect to $t$ to obtain
\begin{equation}
    \dot{s}_{k,m}(t) = \left[1 - \rho_k(0)\right]\left[ \frac{m}{\phi(t)} - \frac{k - m}{1 - \phi(t)}\right]B_{k,m}(\phi(t))\dot{\phi}(t) \,\, \,\, \text{for } \,\, m < k\sigma_k \,.
    \label{eq:skm_red1}
\end{equation}
We then substitute \cref{eq:skm_ansatz} {into} 
\cref{eq:skm} for $\dot{s}_{k,m}$ to obtain
\begin{equation}
    \dot{s}_{k,m} = \alpha\left[1 - \rho_k(0)\right]\left[(k - m + 1)B_{k,m - 1}(\phi) - (k - m)B_{k,m}(\phi)\right] \,\, \,\, \text{for } \,\, m < k\sigma_k \,.
    \label{eq:skm_red2}
\end{equation}
We then equate the right-hand sides of \cref{eq:skm_red1} and \cref{eq:skm_red2} and use
\begin{equation}
    B_{k,m - 1}(\phi) = \frac{1 - \phi}{\phi}\frac{m}{k - m + 1}B_{k,m}(\phi) \,\,\,\,  \text{for } \,\, m \in \{1,2,\ldots,k\} 
\end{equation}
to obtain
\begin{equation}
    \dot{\phi} = \alpha(1 - \phi) \,\, \,\, \text{for } \,\, m < k\sigma_k \,.
    \label{eq:phi_dot_alpha}
\end{equation}

We proceed analogously for the group dynamics. Differentiating \cref{eq:fni_ansatz} with respect to $t$ yields
\begin{equation}
    \dot{f}_{{n},i} = \left[ \frac{i}{\theta} - \frac{n - i}{1 - \theta}\right]B_{n,i}(\theta)\dot{\theta} \, \,\,\, \text{for }\,\, i < n {\tau_n} \,.
    \label{eq:fni_red1}
\end{equation}
We then substitute \cref{eq:fni_ansatz} {into} 
\cref{eq:fni} to obtain
\begin{equation}
    \dot{f}_{n,i} = \left[ \frac{i(1 - \theta)-\theta(n - i)}{\theta} \right]B_{n,i}(\theta)\eta \,\,\,\, \text{for }\,\, i < n{\tau_n} \,.
    \label{eq:fni_red2}
\end{equation}
{To obtain \cref{eq:fni_red2} from \cref{eq:fni_ansatz} and \cref{eq:fni}, we also use the fact that}
\begin{equation}
    B_{n,i - 1}(\theta) = \frac{1 - \theta}{\theta}\frac{i}{n - i + 1}B_{n,i}(\theta) \,\,\,\, \text{for }\,\, i\in \{1,2,\ldots,n\} \,.
\end{equation}
We equate \cref{eq:fni_red1} and \cref{eq:fni_red2} and simplify to obtain
\begin{equation}
    \dot{\theta} = \eta(1 - \theta) \,\,\,\, \text{for }\,\, i < n{\tau_n} \,.
    \label{eq:theta_dot_eta}
\end{equation}

Thus far, we have {obtained} expressions for $\dot{\theta}(t)$ and $\dot{\phi}(t)$ in terms of $\theta(t)$, $\phi(t)$, $\eta(t)$, and $\alpha(t)$. Because $\eta(t)$ and $\alpha(t)$ depend on $f_{n,i}$ and $s_{k,m}$, we also need to obtain expressions for $\eta(t)$ and 
$\alpha(t)$ in terms of $\rho(t)$, $\theta(t)$, and $\phi(t)${. In other words, we want to remove the dependence on $f_{n,i}$ and $s_{k,m}$ from the $(\dot{\theta}(t), \dot{\phi}(t)$) system.} We substitute \cref{eq:skm} into the expression for $\frac{d}{dt}\left[ \sum_{k}g_k \sum_m(k - m)s_{k,m} \right]$ to obtain
\begin{equation}
\begin{split}
    \frac{d}{dt}\left[ \sum_{k = 0}^{{\infty}}g_k \sum_{m = 0}^{{k}}(k - m)s_{k,m} \right]
    	&= -\sum_{k = 0}^{{\infty}} g_k \sum_{{{k\geq}}m\geq k\sigma_k} (k - m)s_{k,m}\\
    	&\quad + \sum_{k = 0}^{{\infty}} g_k \sum_{m = 0}^{{k}} (k - m)(k - m + 1)\alpha s_{k,m - 1}\\
    	&\quad -\sum_{k = 0}^{{\infty}} g_k \sum_{m = 0}^{{k}} (k - m)^2 \alpha s_{k,m} \,.
    \label{eq:eta_to_theta_phi}
    \end{split}
\end{equation}
Using the definition of $\eta(t)$ in \cref{eq:eta}, we write the first term on the right-hand side of \cref{eq:eta_to_theta_phi} in terms of $\eta$ to obtain the first term on the right-hand side of \cref{eq:alpha_eta_to_get_eta}. 
We then use the resulting expression to reduce the right-hand side of  \cref{eq:eta_to_theta_phi} to a single term. We thereby obtain
\begin{equation}
    \begin{split}
        \frac{d}{dt}\left[ \sum_{k = 0}^{{\infty}}g_k \sum_{m = 0}^{{k}}(k - m)s_{k,m} \right]
    &= -\eta\sum_{k = 0}^{{\infty}} g_k \sum_{m = 0}^{{k}} (k - m)s_{k,m}
    - \alpha\sum_{k = 0}^{{\infty}} g_k \sum_{m = 0}^{{k}} (k - m)s_{k,m}\\
    &= -(\alpha + \eta)\sum_{k = 0}^{{\infty}} g_k \sum_{m = 0}^{{k}} (k - m)s_{k,m} \,,
    \end{split} 
    \label{eq:alpha_eta_to_get_eta}
\end{equation}
which we rewrite as
\begin{equation}
    -(\alpha + \eta) = \frac{\frac{d}{dt}\left[ \sum_{k = 0}^{{\infty}}g_k \sum_{m = 0}^{{k}}(k - m)s_{k,m} \right]}{\sum_{k = 0}^{{\infty}}g_k \sum_{m = 0}^{{k}}(k - m)s_{k,m}} = \frac{d}{dt}\left[\ln\left( \sum_{k = 0}^{{\infty}}g_k \sum_{m = 0}^{{k}}(k - m)s_{k,m}\right)\right] \,.
    \label{eq:alpha_eta_to_get_eta2}
\end{equation}

From \cref{eq:phi_dot_alpha} and \cref{eq:theta_dot_eta}, we obtain  
\begin{align}
    \alpha &= \frac{\dot{\phi}}{1 - \phi} = -\frac{d}{dt}\left[\ln(1 - \phi)\right] \,,     \label{eq:phi_dot_alpha2} \\
    \eta &= \frac{\dot{\theta}}{1-\theta} = -\frac{d}{dt}\left[\ln(1 - \theta)\right] \,.
    \label{eq:theta_dot_eta2}
\end{align}
We then combine \cref{eq:alpha_eta_to_get_eta2}, \cref{eq:phi_dot_alpha2}, and \cref{eq:theta_dot_eta2} to obtain
\begin{equation}
    \frac{d}{dt}\left[ \ln((1 - \theta)(1 - \phi)) \right] = \frac{d}{dt}\left[\ln\left( \sum_{k = 0}^{{\infty}}g_k \sum_{m = 0}^{{k}}(k - m)s_{k,m}\right)\right] \,,
\end{equation}
which yields
\begin{equation}
    c_1(1 - \theta)(1 - \phi) = \sum_{k = 0}^{{\infty}}g_k \sum_{m = 0}^{{k}}(k - m)s_{k,m}
    \label{eq:c1_eq}
\end{equation}
by integrating and rearranging. 

Using a similar sequence of arguments, one can show that 
\begin{equation}     \label{eq:res_alph_theta_phi}
    c_2(1 - \theta)(1 - \phi) = \sum_{n = 0}^{{\infty}} p_n \sum_{i = 0}^{{n}} (n - i)f_{n,i} \,.
\end{equation}

We determine the constants $c_1$ and $c_2$ from the initial conditions $\rho(0)$, $\theta(0)$, and $\phi(0)$, the degree distribution ${g_k}$, and the {group-size}
distribution ${p_n}$. The equations for $c_1$ and $c_2$ are
\begin{align}
\label{eq:c1}
    c_1 &= \frac{\sum_{k = 0}^{{\infty}} g_k \sum_{m = 0}^{{k}} (k - m)s_{k,m}(0)}{(1 - \theta(0))(1 - \phi(0))} \,,\\
    c_2 &= \frac{\sum_{n = 0}^{{\infty}} p_n \sum_{i = 0}^{{n}} (n - i) f_{n,i}(0)}{(1 - \theta(0))(1 - \phi(0))}\, . \label{eq:c2}
\end{align}
We want to write Eqs.~\cref{eq:phi_dot_alpha} and \cref{eq:theta_dot_eta} in a form that is independent of $\alpha$ and $\eta$. To do this, we write
\begin{equation}
    \alpha = \frac{\sum_{n = 0}^{{\infty}} p_n\sum_i (n - i) f_{n,i} - \sum_{n = 0}^{{\infty}} p_n\sum_{0 \leq i < n{\tau_n}} (n - i) f_{n,i}}{\sum_{n = 0}^{{\infty}} p_n\sum_i (n - i) f_{n,i}} \,,
\end{equation}
which allows us to use \cref{eq:res_alph_theta_phi} and the ansatz \cref{eq:fni_ansatz} to obtain
\begin{equation}
    \alpha = \frac{c_2(1 - \theta)(1 - \phi) - \sum_{n = 0}^{{\infty}} p_n \sum_{0 \leq i < n{\tau_n}}(n - i)B_{n,i}(\theta)}{c_2(1 - \theta)(1 - \phi)} \,.
    \label{eq:alpha_ov}
\end{equation}
Similarly, {we obtain}
\begin{equation}
    \eta = \frac{c_1(1 - \theta)(1 - \phi) - \sum_{k = 0}^{{\infty}} g_k \sum_{0 \leq m < k\sigma_k}\left[1 - \rho_k(0)\right](k - m)B_{k,m}(\phi)}{c_1(1 - \theta)(1 - \phi)} \,.
    \label{eq:eta_ov}
\end{equation}
We substitute the right-hand side of \cref{eq:alpha_ov} into \cref{eq:phi_dot_alpha} to eliminate $\alpha$ and substitute the right-hand side of \cref{eq:eta_ov} into \cref{eq:theta_dot_eta} to eliminate $\eta$. We thereby obtain 
\begin{align}
    \dot{\phi} &= \begin{cases}\label{eq:phi_dot_2nd}
   			 \frac{c_2(1 - \theta)(1 - \phi) - \sum_{n = 0}^{{\infty}} p_n \sum_{0 \leq i < n{\tau_n}}(n-i)B_{n,i}(\theta)}{c_2(1 - \theta)}& \text{if } \theta < 1 \\
  				  0&\text{otherwise} \,,
			\end{cases} \\
    \dot{\theta} &=
    	\begin{cases}\label{eq:theta_dot_2nd}
   		 \frac{c_1(1 - \theta)(1 - \phi) - \sum_{k = 0}^{{\infty}} g_k \sum_{0 \leq m < k\sigma_k}\left[1 - \rho_k(0)\right](k - m)B_{k,m}(\phi)}{c_1(1 - \phi)} &\text{if }\phi < 1 \\
   		 0&\text{otherwise} \,,
    \end{cases}
\end{align}
where $c_1$ and $c_2$ are given by \cref{eq:c1} and \cref{eq:c2}, respectively. 


{To} obtain $\dot{\rho}$, we write $\rho(t)$ in the form
\begin{equation}
    \rho(t) = 1 - \sum_{k = 0}^{{\infty}} g_k \sum_{m = 0}^{{k}} s_{k,m} 
\end{equation}
and differentiate with respect to $t$ to obtain
\begin{equation} \label{above}
    \dot{\rho} = -\sum_{k = 0}^{{\infty}} g_k \sum_{m = 0}^{{k}} \dot{s}_{k,m} \,.
\end{equation}
We then substitute \cref{eq:skm} into 
\cref{above} to yield
\begin{equation}
    \dot{\rho} = -\left[ -\sum_{k = 0}^{{\infty}} g_k \sum_{k\sigma_k \leq m \leq k}s_{k,m}+\alpha\sum_{k = 0}^{{\infty}} g_k \sum_{m = 1}^{{k}} (k - m + 1)s_{k,m - 1} - \alpha\sum_{k = 0}^{{\infty}} g_k \sum_{m = 0}^{{k}} (k - m)s_{k,m}\right] \,,
\end{equation}
where the first term on the right-hand side arises from
$\gamma(k,m)$ in \cref{eq:skm} equaling 1 for $m \geq k \sigma_k$ and the last two terms on the right-hand side telescope to $0$. We thereby obtain
\begin{equation}
    \dot{\rho} = \sum_{k = 0}^{{\infty}} g_k \sum_{k\sigma_k \leq m \leq k}s_{k,m} \,.
    \label{eq:rho_dot_prelim}
\end{equation}
One can further rewrite 
\cref{eq:rho_dot_prelim} as
\begin{equation}
\begin{split}
    \dot{\rho} &= \sum_{k = 0}^{{\infty}} g_k \sum_{m = 0}^{k}s_{k,m} - \sum_{k = 0}^{{\infty}} g_k \sum_{0 \leq m < k\sigma_k}s_{k,m} \\
    		&= (1 - \rho) -\sum_{k = 0}^{{\infty}} g_k [1 - \rho_k(0)]\sum_{0 \leq m < k\sigma_k}B_{k,m}(\phi) \,.
    \end{split}
    \label{eq:rho_dot_2nd}
\end{equation}

Now that we have expressions for $\dot{\phi}$, $\dot{\theta}$, and $\dot{\rho}$ in terms of $\phi$, $\theta$, and $\rho$, the last step is to determine the constants $c_1$ and $c_2$ (see Eqs.~\cref{eq:c1} and \cref{eq:c2}) that appear in the equations for $\dot{\theta}$ (see \cref{eq:theta_dot_2nd}) and $\dot{\phi}$ (see \cref{eq:phi_dot_2nd}), respectively, from {our} initial conditions. 
 Because we seed the nodes uniformly at random (and hence independently of their degrees), the probability that a uniformly randomly selected node {in}
an inactive group is initially active is $\theta(0) = \rho_0$ and the fraction of degree-$k$ nodes that are initially active {is} $\rho_k(0) =
\rho_0$. {We obtain the value of $\phi_0$ directly from \cref{eq:phi_0}.}


\section{Cascade Condition} \label{sec:cascade_condition}

A ``cascade condition" indicates whether or not a system experiences a global cascade in a given situation~\cite{lehmann2018}. For example, perhaps many individuals in a social network adopt the same type of technology (e.g., a smartphone)~\cite{watts2002simple}, many institutions default in a financial system~\cite{gleeson2013systemic, elliott2014financial}, or many components of a power grid {fail}
and thereby lead to a blackout~\cite{schafer2018dynamically, chang2011performance}.

{Our double-threshold hypergraph WTM has} a global cascade if $\rho(t) \not\rightarrow 0$ as $t \to \infty$. To derive an approximate cascade condition {to describe when there is a global cascade}, we linearize the system of equations for $(\dot{\theta}(t),\dot{\phi}(t))$ {({see Eqs.~(\ref{eq:theta_dot}, \ref{eq:phi_dot})})}, calculate the Jacobian matrix for the linearized system around $\theta = \phi = 0$, and determine {when} at least one of its eigenvalues is positive. {The} cascade condition is approximate because we derive it from a linearization of the $(\dot{\theta}, \dot{\phi})$ system, which is an exact reduction of the full AME system \cref{eq:fni}--\cref{eq:alpha}. Because we linearize about $0$, this cascade condition is most accurate when $\rho_0$ is small (and especially when $\rho_0 \approx 0$). We consider the ($\dot{\theta},\dot{\phi}$) system instead of the full $(\dot{\rho}, \dot{\theta}, \dot{\phi})$ system ({see Eqs.~\cref{eq:rho_dot}--\cref{eq:phi_dot}}{)} because $\dot{\theta}$ and $\dot{\phi}$ are independent of $\rho$. Including the expression for $\dot{\rho}$ gives an additional eigenvalue with value $-1$. 

For the ($\dot{\theta},\dot{\phi}$) system, the Jacobian matrix around $\theta = \phi = 0$ is
\begin{equation}
	J =  \begin{pmatrix}
        			-1 & \frac{\partial \dot{\theta}}{\partial \phi}\bigg\rvert_{(0,0)} \\
        			\frac{\partial \dot{\phi}}{\partial \theta}\bigg\rvert_{(0,0)} & -1
    	\end{pmatrix} \,,
\end{equation}
which yields the eigenvalues
\begin{equation}
    \lambda_{1,2} = -1 \pm \sqrt{\frac{\partial \dot{\theta}}{\partial \phi}\bigg\rvert_{(0,0)}\frac{\partial \dot{\phi}}{\partial \theta}\bigg\rvert_{(0,0)}} \, \, .
\end{equation}
At least one of these eigenvalues is positive if and only if $\frac{\partial \dot{\theta}}{\partial \phi}\bigg\rvert_{(0,0)}\frac{\partial \dot{\phi}}{\partial \theta}\bigg\rvert_{(0,0)} > 1$, so there is a global cascade if
\begin{equation}
    \frac{\partial \dot{\theta}}{\partial \phi}\bigg\rvert_{(0,0)}\frac{\partial \dot{\phi}}{\partial \theta}\bigg\rvert_{(0,0)} > 1 \,,
    \label{eq:cascade_condition}
\end{equation}
where
\begin{align}
    \frac{\partial \dot{\theta}}{\partial \phi} &= \frac{\sum_{\{k | k\sigma_k \leq 1\}}g_k (1 - \rho_k (0))k(k - 1)}{c_1} \,, \\
    \frac{\partial \dot{\phi}}{\partial \theta} &= \frac{\sum_{\{n | n{\tau_n}\leq 1\}} p_n n(n - 1)}{c_2} \,.\label{eq:dphidot_dtheta}
\end{align}

\begin{figure}[h]
    \centering
    \subfloat[]{\includegraphics[width=0.49\linewidth]{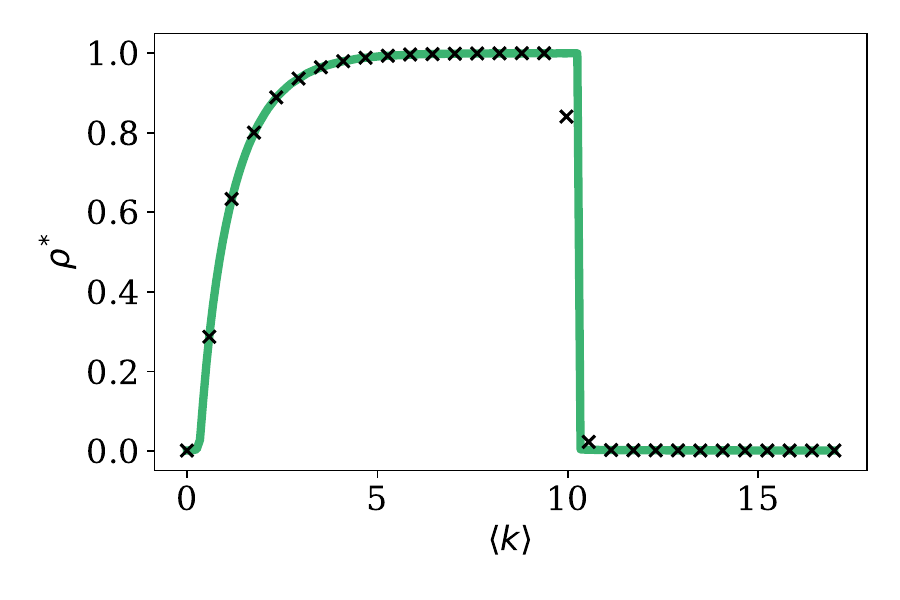}}
    \subfloat[]{\includegraphics[width=0.49\linewidth]{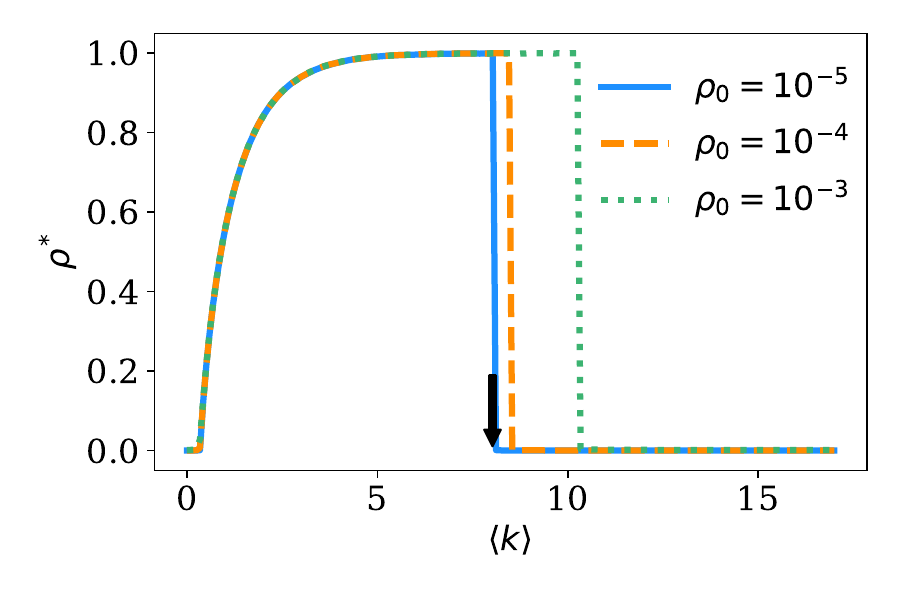}}
    \caption{(a) The steady-state fraction $\rho^{*}$ of active nodes {in the reduced AME system}~{\cref{eq:rho_dot}--\cref{eq:phi_dot}} for an initially active node fraction $\rho_0 = 10^{-3}$, degree distribution $g_k \sim \mathrm{Pois}(\langle k \rangle)$, 
    {group-size}
    distribution $p_n\sim\mathrm{Pois}(3)$, node threshold $\sigma_k = 0.18$, and group threshold ${\tau_n} = 0.1$. The black markers indicate the mean steady-state densities of active nodes from 100 WTM simulations on a single 50,000-node configuration-model hypergraph. In our numerical calculations, we suppose that $\rho(t)$ has attained a steady state by time $t = 100$. (b) The steady-state fraction of active nodes {in the reduced AME system} for initially active node fractions $\rho_0 = 10^{-5}$ (solid blue curve), $\rho_0 = 10^{-4}$ (dashed orange curve), and $\rho_0 = 10^{-3}$ (dotted green curve). The black arrow points to the critical value of $\langle k \rangle$ that we calculate from the linearization of the reduced AME system~\cref{eq:rho_dot}--\cref{eq:phi_dot}. For each of these values of $\rho_0$, the approximate critical degree from the linearization of the $(\dot{\theta}, \dot{\phi})$ system is $\langle k \rangle\approx 8.02$ {(with different values {of the critical $\langle k \rangle$} in the third decimal place for the three values of $\rho_0$)}{.}
 }   
    \label{fig:cc_nodes}
\end{figure}

\begin{figure}
    \centering
    \subfloat[]{\includegraphics[width=0.49\linewidth]{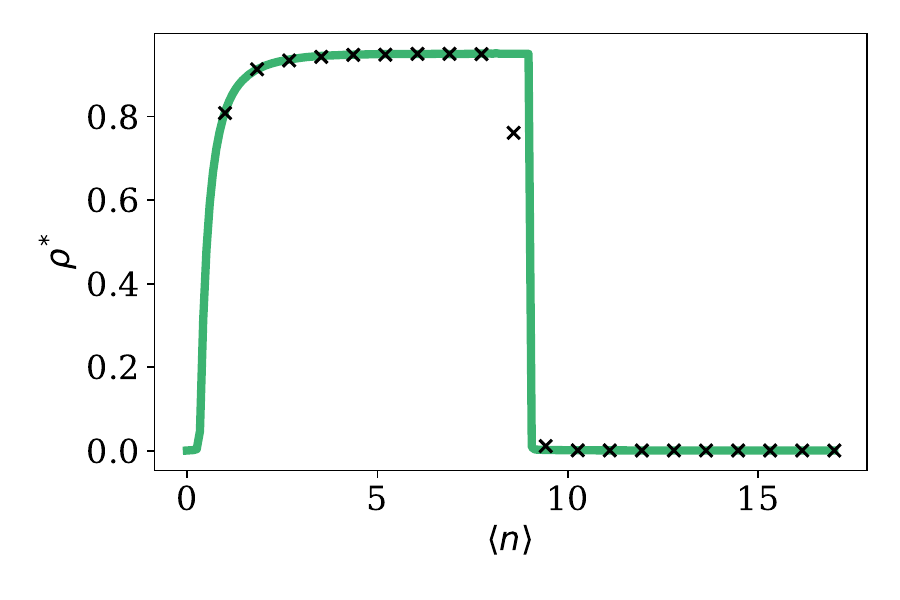}}
    \subfloat[]{\includegraphics[width=0.49\linewidth]{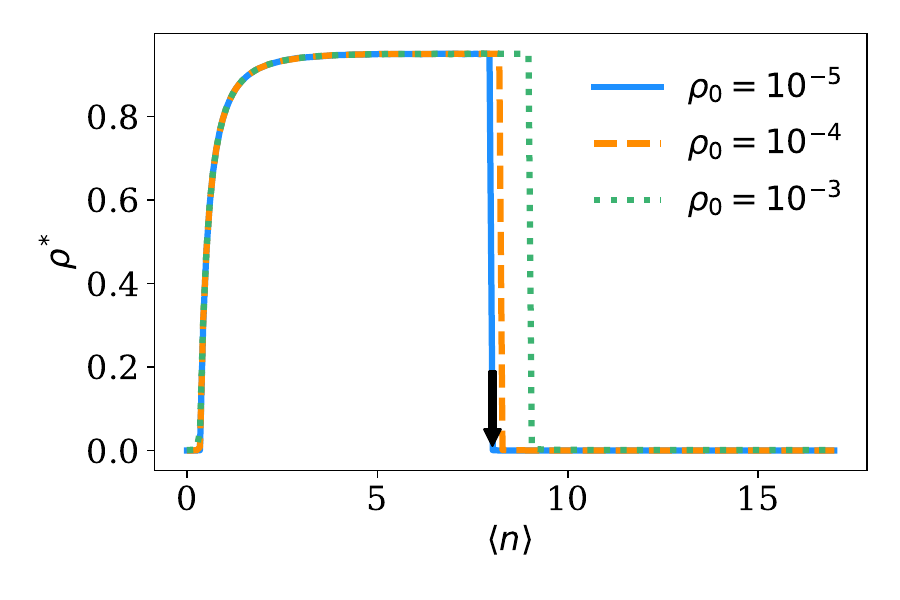}}
    \caption{(a) The steady-state fraction $\rho^{*}$ of active nodes {in the reduced AME system}~{\cref{eq:rho_dot}--\cref{eq:phi_dot}} for an initially active node fraction $\rho_0 = 10^{-3}$, degree distribution $g_k \sim \mathrm{Pois}(3)$, 
    {group-size} distribution $p_n \sim \mathrm{Pois}(\langle n \rangle)$, node threshold 
    $\sigma_k = 0.1$, and group threshold ${\tau_n} = 0.18$. The black markers indicate the mean steady-state densities of active nodes from 100 WTM simulations on a single 50,000-node configuration-model hypergraph. (b) The steady-state fractions of active nodes {in the reduced AME system} for the degree distribution $g_k\sim\mathrm{Pois}(3)$, {group-size} distribution $p_n \sim \mathrm{Pois}(\langle n \rangle)$, node threshold $\sigma_k = 0.1$, and group threshold ${\tau_n} = 0.18$ for initially active node fractions $\rho_0 = 10^{-5}$ (solid blue curve), $\rho_0 = 10^{-4}$ (dashed orange curve), and $\rho_0 = 10^{-3}$ (dotted green curve). The black arrow points to the critical {value of} $\langle n \rangle$ that we calculate from the linearization of the reduced AME system~\cref{eq:rho_dot}--\cref{eq:phi_dot}. For each of these values for $\rho_0$, the critical hyperedge size is 
     $\langle n \rangle\approx 8.02$ {(with different values {of the critical $\langle n \rangle$} in the third decimal place for the three values of $\rho_0$)}{.}
     }
    \label{fig:cc_edges}
\end{figure}

In \cref{fig:cc_nodes}(a) and \cref{fig:cc_edges}(a), we show the transition between a large expected steady-state fraction of active nodes to a very small steady-state fraction of active nodes as we increase the mean degree $\langle k \rangle$ (see \cref{fig:cc_nodes}(a)) or the mean group size $\langle n \rangle$ (see \cref{fig:cc_edges}(b)) for {group} sizes and degrees that follow Poisson distributions. In \cref{fig:cc_nodes}(b) and \cref{fig:cc_edges}(b), we show the steady-state $\rho^*$ {that we obtain using} the {reduced} AME system~{\cref{eq:rho_dot}--\cref{eq:phi_dot}} {versus} $\langle k \rangle$ and $\langle n \rangle$, respectively, for different initially active node fractions $\rho_0$. The black arrows in these plots mark the values of 
$\langle k \rangle$ and $\langle n \rangle$ from the cascade condition \cref{eq:cascade_condition} {that indicate {very sharp}
transitions between global cascades and very little {contagion spread}.}
{Gleeson and Cahalane~\cite{gleeson2007seed} showed that the analogous transitions are discontinuous in the WTM on ordinary graphs.}
{When} the initial active node fraction $\rho_0 \to 0$, we see in \cref{fig:cc_nodes}(b) and \cref{fig:cc_edges}(b) that the cascade condition \cref{eq:cascade_condition}, which we obtained by linearizing the reduced AME system~\cref{eq:rho_dot}--\cref{eq:phi_dot}, becomes increasingly accurate.


\section{Results for Empirical Networks} \label{sec:empirical} 
%
{We now examine the dynamics of} our continuous-time double-threshold hypergraph WTM on two real-world hypergraphs. {{The purpose of comparing direct numerical simulations of our WTM to AME solutions 
for real-world networks is to give insight into where our theoretical approximations succeed
in real-world networks, which have finitely many nodes and include correlations that the AMEs do not incorporate.}}
These {{real-world}} hypergraphs are {(i)} a French primary-school face-to-face contact network (with 242 nodes and 1188 hyperedges), which was collected by Stehle et al.~\cite{stehle2011high} and adapted to a hypergraph form by St-Onge et al.~\cite{stonge2022}{,} and {(ii)} a DBLP (Digital Bibliography \& Library Project) computer-science coauthorship hypergraph. 
DBLP is an online system that collects information about publications in computer-science journals and conference proceedings.
The DBLP coauthorship network was assembled by Benson et al.~\cite{benson2018simplicial}, but we use the subhypergraph of it that St-Onge et al.~\cite{stonge2022} obtained using a breadth-first search. The full coauthorship network has 1,831,127 nodes and 2,954,518 hyperedges; the examined subhypergraph has 57,501 nodes and 55,204 hyperedges.

\begin{figure}[h]
    \centering
    \subfloat[]{\includegraphics[width = 0.49\textwidth]{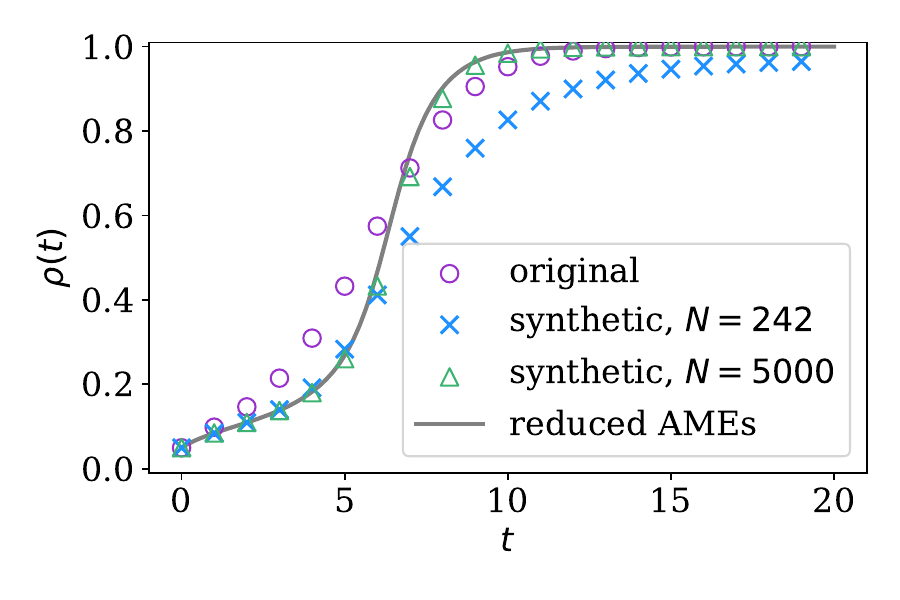}}
    \subfloat[]{\includegraphics[width = 0.49\textwidth]{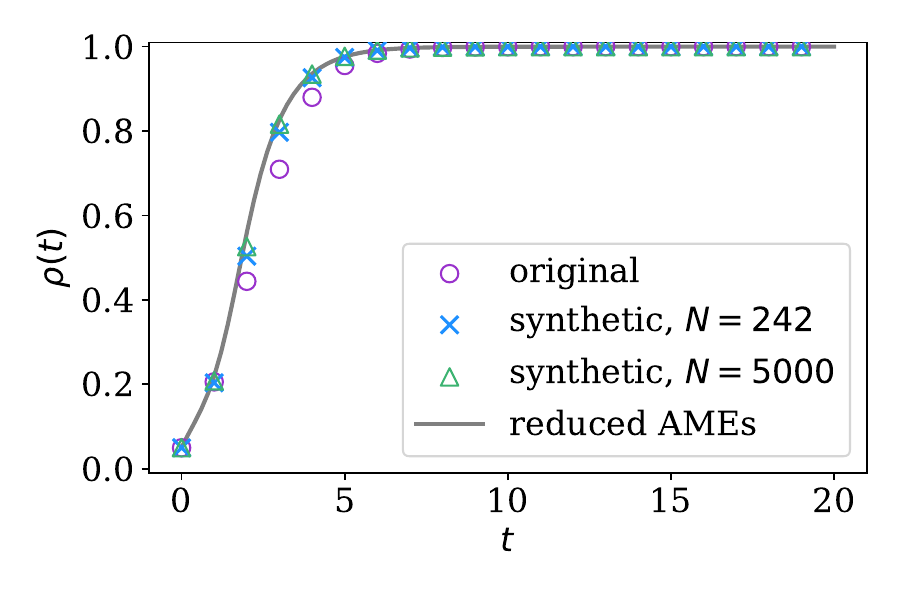}}
   \caption{The fraction $\rho(t)$ of active nodes in our continuous-time double-threshold hypergraph WTM on the French primary school face-to-face contact network~\cite{stehle2011high,stonge2022}. 
   This hypergraph has {242} nodes, 1188 hyperedges, {a mean degree of $\langle k\rangle \approx 11.79$, a mean hyperedge (i.e., group) size of} $\langle n\rangle \approx 2.4$, a maximum degree of 32, and a maximum group size of 5. 
   We show the results of computations with (a) 
   {initially} active node fraction $\rho_0 = 0.05$, node threshold $\sigma_k = 0.25$, and group threshold ${\tau_n} = 0.3$ and (b)
   {initially} active node fraction $\rho_0 = 0.05$, node threshold $\sigma_k = 0.15$, and group threshold ${\tau_n} = 0.2$. 
      The solid gray curves are solutions of the reduced AME system~\cref{eq:rho_dot}--\cref{eq:phi_dot},
    and the purple circles are mean values of $\rho(t)$ from 500 simulations of the double-threshold hypergraph WTM on the original contact hypergraph. The blue crosses are means of simulations of the 
   double-threshold hypergraph WTM on 500 different 242-node configuration-model hypergraphs that we generate with the same degree distribution $\{g_k\}$ and group-size distribution $\{p_{n}\}$ as in the original contact hypergraph, and the green triangles are the means of simulations of the double-threshold hypergraph WTM on 500 different 5000-node configuration-model hypergraphs with the same {$\{g_k\}$ and $\{p_n\}$} as in the original contact hypergraph. 
   }
    \label{fig:school_example2}
\end{figure}

In \cref{fig:school_example2}, we compare the results of the reduced AME system {\cref{eq:rho_dot}--\cref{eq:phi_dot}}
(solid gray curve) to simulations of the continuous-time double-threshold hypergraph WTM on the primary-school face-to-face contact network. For this comparison, we input the degree distribution and the {group-size}
distribution of the empirical network into the reduced AME system to generate our results. The initially active node fraction is $\rho_0 = 0.05$. In \cref{fig:school_example2}(a), the node threshold is $\sigma_k = 0.25$ for all nodes and the group threshold is ${\tau_n} = 0.3$ for all hyperedges. In \cref{fig:school_example2}(b), which {depicts} a faster-growing contagion than the one in \cref{fig:school_example2}(a), the node threshold is $\sigma_k = 0.15$ for all nodes and the group threshold is ${\tau_n} = 0.2$ for all hyperedges. We obtain good agreement between the reduced-AME results and the WTM simulations, although the agreement is not perfect. We believe that the discrepancy between the reduced-AME results and direct simulations {of the double-threshold hypergraph WTM} arise from the small network size (of 242 nodes) {and from correlations} between the degrees of different nodes, {correlations} between the sizes of different hyperedges, and {correlations} between {node degrees and hyperedge {(i.e., group)} sizes.} 
Our full and reduced AME systems do not account for {any} of these correlations. 


{To} explore the effects of correlations, we fix the number of nodes to match the number in the {primary-school face-to-face contact network}
and we generate configuration-model hypergraphs with the same degree distribution $\{g_k\}$ and the same {group-size} 
distribution $\{p_n\}$ (blue crosses). We thereby account for the impact of these correlations in the primary-school face-to-face contact network.
In \cref{fig:school_example2}(a), we observe a very close match between the reduced-AME results and {WTM} simulations for small {active-node fractions} $\rho(t)$ but a much weaker match for large $\rho(t)$. In \cref{fig:school_example2}(b), which has 
{smaller} values of the node threshold and group threshold, we obtain an almost perfect match between the reduced-AME results and WTM simulations. It is evident from \cref{fig:school_example2}(a) that the small network size also impacts the {quality} of the reduced-AME results. Therefore, we increase the network size to 5000 nodes and generate 500 different configuration-model hypergraphs using {$\{g_k\}$ and $\{p_n\}$} from the primary-school face-to-face contact network (green triangles).
As we can see in both panels of \cref{fig:school_example2}, we now obtain excellent agreement between the reduced-AME results and direct simulations of the double-threshold hypergraph WTM. We thus conclude that the discrepancy between the reduced AME system {\cref{eq:rho_dot}--\cref{eq:phi_dot}} and WTM simulations on the {primary-school face-to-face contact network}
arises both from finite-size effects and from correlations in the real-world hypergraph that are not captured by our full or reduced AME systems.

\begin{figure}[h]
    \centering\subfloat[]{\includegraphics[width = 0.49\textwidth]{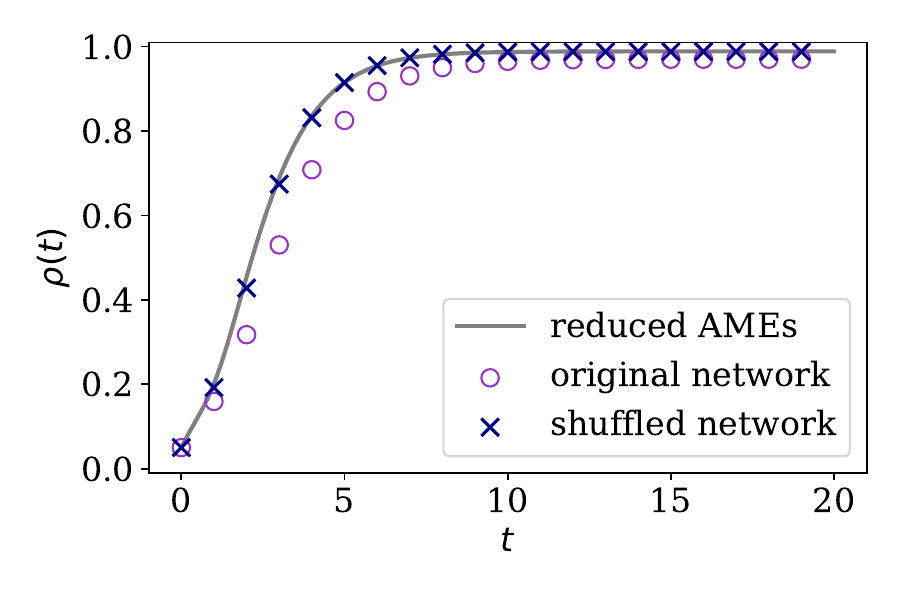}}
    \centering\subfloat[]{\includegraphics[width = 0.49\textwidth]{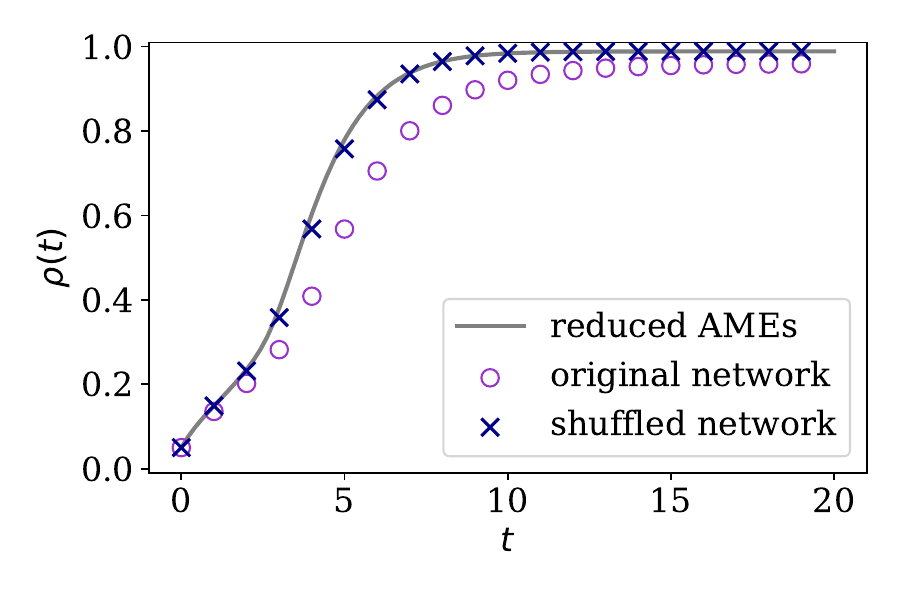}}
   \caption{The fraction $\rho(t)$ of active nodes in our continuous-time double-threshold hypergraph WTM on a subhypergraph of the DBLP computer-science coauthorship network.
   This subhypergraph has {57,501} nodes, 55,204 hyperedges, {a mean degree of} $\langle k \rangle \approx 3.75$, {a mean hyperedge (i.e., group) size of} $\langle n \rangle \approx 3.90$, a maximum degree of 903, and a maximum group size of 25. We show the results of computations with (a) 
   {initially} active node fraction $\rho_0 = 0.05$, node threshold $\sigma_k = 0.15$, group threshold ${\tau_n} = 0.2$ and (b) 
   {initially} active node fraction $\rho_0 = 0.05$, node threshold $\sigma_k = 0.2$, and group threshold ${\tau_n} = 0.25$. The solid gray curves are solutions of the reduced AME system~\cref{eq:rho_dot}--\cref{eq:phi_dot}, and the purple circles are 
   {are mean values of $\rho(t)$ from} 500 simulations of the double-threshold hypergraph WTM on the coauthorship subhypergraph. The blue crosses are means of 500 WTM simulations on a single shuffled coauthorship subhypergraph, in which we shuffle the nodes uniformly at random among the hyperedges.
    }
    \label{fig:coauthor_example1}
\end{figure}

In \cref{fig:coauthor_example1}, we compare solutions of the reduced AME system~\cref{eq:rho_dot}--\cref{eq:phi_dot} on {a subhypergraph of} the DBLP computer-science coauthorship network \cite{benson2018simplicial, stonge2022} to simulations of the double-threshold hypergraph WTM on it for two different sets of threshold distributions. We show results for {node} threshold $\sigma_k = 0.15$ {(for all nodes)} and group threshold ${\tau_n} = 0.2$ {(for all hyperedges)} in \cref{fig:coauthor_example1}(a), and we show results for 
{node} threshold $\sigma_k = 0.2$ and group threshold {${\tau_n} = 0.25$} in \cref{fig:coauthor_example1}(b). In both cases, the initially active fraction of nodes is $\rho_0 = 0.05$. The reduced AME system yields results that resemble those of the {WTM} simulations, but there are some correlations that the AME system does not capture. The coauthorship {subhypergraph} has 57,501 nodes, so we do not expect finite-size effects to {produce}
any significant {discrepancies between AME results and direct simulationrs of the WTM}. To confirm this expectation, we shuffle the nodes among the hyperedges. In our shuffling procedure, we preserve the node degrees and the {hyperedge} 
sizes, but we uniformly randomly assign the nodes to hyperedges. We then simulate our continuous-time double-threshold hypergraph WTM on the shuffled network and find extremely strong agreement between these simulations and the reduced-AME results. For this example, we shuffle nodes among the hyperedges instead of generating a synthetic hypergraph for two reasons. First, the DBLP coauthorship {subhypergraph}
is already large, so we do not need to enlarge it to account for finite-size effects. Second, the DBLP coauthorship {subhypergraph}
has some nodes with very large degrees (with $k_{\text{max}} = 903$). 
Due to the large degrees, it takes a long time to generate a configuration model from the joint distribution of degrees and {group}
sizes. 
{A configuration-model hypergraph}
requires the sum of the degrees to equal the sum of the {group}
sizes, and satisfying this constraint poses computational difficulties. When we introduce a large group (i.e., a hyperedge that is attached to many nodes), the sum of the {group}
sizes increases by a large number and this sum is like to exceed the sum of the degrees. If this occurs, we resample both the degree sequence and the {group-size}
sequence. This situation can occur repeatedly, and then it takes a long time to generate a desired hypergraph. 


\section{Conclusions and Discussion}
\label{sec:discussion}

Social systems include both dyadic and polyadic interactions, and it is thus important to generalize models of social dynamics and analytical approaches to analyze such models from graphs to hypergraphs. One foundational model of social dynamics is the Watts threshold model (WTM), which describes a simplistic social contagion and has been studied by many researchers for more than two decades.

In the present paper, we derived a system of approximate master equations (AMEs) that accurately describe a continuous-time double-threshold WTM on hypergraphs. We showed that this AME system is accurate both at modeling the expected steady-state dynamics and at approximating the time-dependent fraction of active nodes. The accuracy of this high-dimensional AME system is {one of its key benefits}, but two key drawbacks 
{are} that it is more difficult to analyze and {much slower computationally}
to solve numerically than a mean-field approximation of the double-threshold hypergraph WTM. To overcome these drawbacks, we reduced this high-dimensional AME system using two ansatzes (which are similar to those that Gleeson \cite{gleeson2013} employed for the {standard}
dyadic WTM) to obtain a three-dimensional AME system that retains the high accuracy of the full AME system. This is an exact reduction of the full AME system.

Using the low-dimensional reduced AME system, 
{we} derived an approximate cascade condition, which allows one to determine whether or not our continuous-time double-threshold WTM experiences global cascades. In our numerical computations, we observed that {this}
cascade condition is accurate when the initially active fraction of nodes is small but that it is not accurate for {large} initially active fractions of nodes. We derived our approximate cascade condition by linearizing the three-dimensional reduced AME system around the origin, and we expect that one can derive a more accurate cascade condition by incorporating nonlinear terms.

We also examined the performance of our reduced AME system on real-world hypergraphs from a primary-school face-to-face contact network \cite{stehle2011high, stonge2022} and a subset of a DBLP computer-science coauthorship network \cite{benson2018simplicial, stonge2022}. {Although} the reduced AME {system} performs reasonably well for {these real-world networks,}
we 
{observed} that it does not account for finite-size effects or for structural correlations (e.g., between the degrees of nodes in the same group, between the sizes of groups that share nodes, and between the node degrees and the group sizes). It is worthwhile to generalize our AME systems to account for 
{structural} correlations. 
{{
A few years ago, Peralta and Toral~\cite{peralta2020binary} examined the impact of system size on AME descriptions of binary-state dynamics in several models with dyadic interactions, and it will be worthwhile to conduct analogous 
investigations of finite-size effects 
in AME descriptions of WTMs
with polyadic interactions.}}


\section*{Acknowledgements}

{We thank James Gleeson and Laurent H\'{e}bert-Dufresne for useful discussions, Juwon Kim for many helpful comments on our manuscript, and {three} anonymous referees for useful feedback.}
K.-I.G. thanks the Department of Mathematics at UCLA for hosting him during part of this work {and acknowledges the support of a National Research Foundation of Korea (NRF) grant (No. RS-2025-00558837) funded by the Korea government (MSIT).}


\appendix



\section{Comparison {of our analysis} with Chen et al.~\cite{chen2025simple}}
\label{sec:chen}
%
%
%
In this appendix, we show that the reduced AME system {\cref{eq:rho_dot}--\cref{eq:phi_dot}},
which approximates the {full AME system {\cref{eq:fni}--\cref{eq:alpha}} {of our}
continuous-time double-threshold hypergraph WTM,}
agrees very well with {Chen et al.'s discrete-time method~\cite{chen2025simple}}, which approximates a discrete-time hypergraph double-threshold WTM {and yields}
the self-consistent equation~\cref{eq:chen_steady_state} below. 


{In} \cref{fig:comparison_with_chen}, we plot the steady-state fraction $\rho^*$ of active nodes from both our reduced AME system  {\cref{eq:rho_dot}--\cref{eq:phi_dot}} and the method of Chen et al. for configuration-model hypergraphs with Poisson-distributed node degrees and group sizes {for}
different mean degrees and mean group sizes. Our {reduced-AME} results are indistinguishable from the results of Chen et al.'s discrete-time method. To generate the dashed green curve in \cref{fig:comparison_with_chen} from Chen et al.'s method, one solves Eqs.~(14)--(20) of \cite{chen2025simple} for $\rho^*$ {to obtain}
\begin{align} \label{eq:chen_steady_state}
    \rho^* = \rho_0 + (1 - \rho_0)\sum_{k = 1}^{\infty}g_k\sum_{m = 0}^{k}\binom{k}{m}u_{\infty}^m (1 - u_\infty)^{k - m}\gamma(k,m) \,,
\end{align}
where {$u_\infty \in [0,1]$} 
 is the smallest fixed point of
\begin{equation}
    u_{n + 2} = g(\rho_0 + (1 - \rho_0)f(u_n)) \,,
    \label{eq:u_np2}
\end{equation}
{with
\begin{align}
    g(\omega) &= \sum_{n = 1}^{\infty}\frac{np_n}{\langle n \rangle}\sum_{i = 0}^{n - 1}\binom{n - 1}{i}\omega^i(1 - \omega)^{n - 1 - i}\beta(n,i) \,, \label{eq:g_w}  \\
   	 f(u) &= \sum_{k = 1}^\infty \frac{k g_k}{\langle k \rangle}\sum_{m = 0}^{k - 1}\binom{k - 1}{m}u^m (1 - u)^{k - 1 - m}\gamma(k,m) \,, \label{eq:f_u}
\end{align}
where $\omega$ is the probability that a uniformly random node that one reaches via a uniformly random hyperedge is active and $u$ is the probability that a uniformly random hyperedge that one reaches {via}
a uniformly random node is active.} To obtain $u_\infty$, we let $u_0 = \rho_0$ and iterate \cref{eq:u_np2} until $|u_{n + 2} - u_n| < 10^{-5}$. 
We then treat the resulting value of $u_{n + 2}$ as $u_\infty$. The solid {green} curves in \cref{fig:comparison_with_chen} are the same as the solid green curves in \cref{fig:cc_nodes}(a) and \cref{fig:cc_edges}(a), where we take the solution of the reduced AME system {\cref{eq:rho_dot}--\cref{eq:phi_dot}}
 at $t = 100$ as the steady state $\rho^*$. 
 By contrast, using their discrete-time approach, Chen et al.~\cite{chen2025simple} obtained $\rho^*$ by taking the $t \to \infty$ limit of $\rho(t)$. In \cref{fig:comparison_with_chen}, we see that the values of $\rho^*$ from the reduced AME system agree very well with {the values of $\rho^*$} from Chen et al.'s method.

\begin{figure}[h]
    \centering\subfloat[]{\includegraphics[width = 0.49\textwidth]{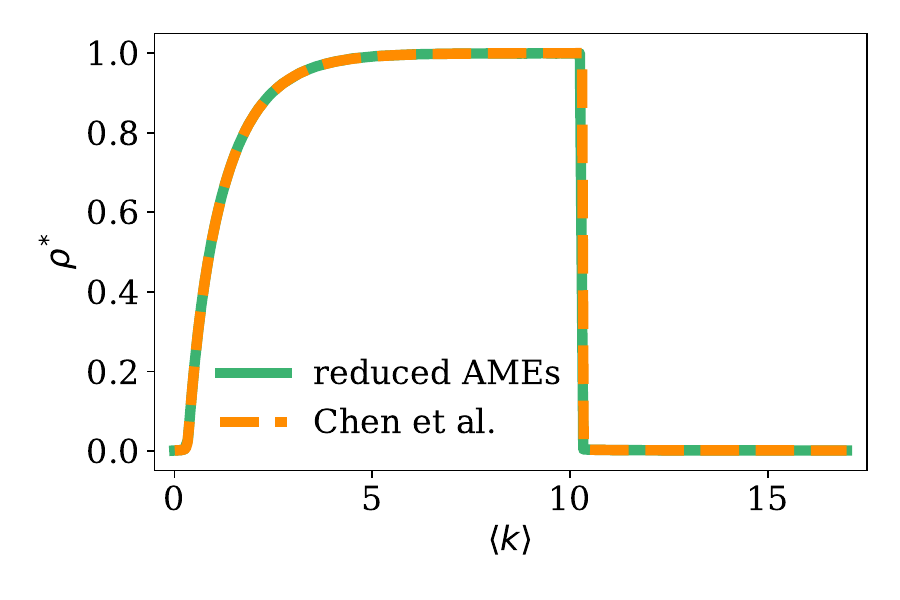}}
    \centering\subfloat[]{\includegraphics[width = 0.49\textwidth]{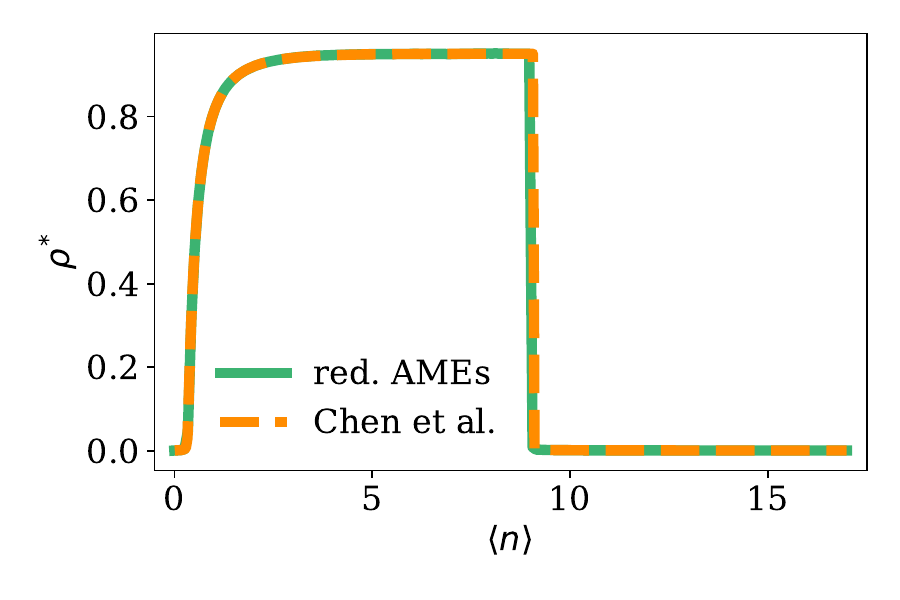}}
   \caption{The steady-state fraction $\rho^*$ of active nodes from the reduced AME system {\cref{eq:rho_dot}--\cref{eq:phi_dot}} (solid green curve) and from the discrete-time system \cref{eq:chen_steady_state}--\cref{eq:f_u} of {Chen et al.~\cite{chen2025simple}} (dashed orange curve) for initially active seed fraction $\rho_0 = 10^{-3}$. 
   We show results of computations on configuration-model hypergraphs with (a) degree distribution $g_k\sim\mathrm{Pois}(\langle k \rangle$), {group-size} distribution $p_n\sim\mathrm{Pois}(3)$, node threshold $\sigma_k = 0.18$, and group threshold ${\tau_n} = 0.1$ and (b) degree distribution $g_k\sim\mathrm{Pois}(3)$, {group-size} distribution $p_n\sim\mathrm{Pois}(\langle n \rangle)$, node threshold $\sigma_k = 0.1$, and group threshold ${\tau_n} = 0.18$.}
   \label{fig:comparison_with_chen}
\end{figure}


\bibliographystyle{siamplain}
\bibliography{refs-v14}

@article{chodrow2020configuration,
  title={Configuration models of random hypergraphs},
  author={Chodrow, Philip S.},
  journal={Journal of Complex Networks},
  volume={8},
  number={3},
  eid={cnaa018},
  year={2020}
}

@article{kim2026,
  title={Optimality in group-driven social dynamics on hypergraphs},
  author={Jihye Kim and Deok-Sun Lee and K.-I. Goh},
  journal={arXiv:2604.17689},
  year={2026}
}

@article{peixoto2026,
  title={Graphs are maximally expressive for higher-order interactions},
  author={Tiago P. Peixoto and Leto Peel and Thilo Gross and Manlio De Domenico},
  journal={arXiv:2602.16927},
  year={2026}
}

@article{pastor2015,
  title = {Epidemic processes in complex networks},
  author = {Pastor-Satorras, Romualdo and Castellano, Claudio and Van Mieghem, Piet and Vespignani, Alessandro},
  journal = {Reviews of Modern Physics},
  volume = {87},
  issue = {3},
  pages = {925--979},
  year = {2015}
}

@book{porter2016,
	Author = {Porter, M. A. and Gleeson, J. P.},
	Publisher = {Springer International Publishing},
	address = {Cham, Switzerland},
	Series = {Frontiers in Applied Dynamical Systems: Reviews and Tutorials},
	Title = {Dynamical Systems on Networks: {A} Tutorial},
	Volume = {4},
	Year = {2016}}

@ARTICLE{2020SciPy-NMeth,
  author  = {Virtanen, Pauli and Gommers, Ralf and Oliphant, Travis E. and
            Haberland, Matt and Reddy, Tyler and Cournapeau, David and
            Burovski, Evgeni and Peterson, Pearu and Weckesser, Warren and
            Bright, Jonathan and {van der Walt}, St{\'e}fan J. and
            Brett, Matthew and Wilson, Joshua and Millman, K. Jarrod and
            Mayorov, Nikolay and Nelson, Andrew R. J. and Jones, Eric and
            Kern, Robert and Larson, Eric and Carey, C J and
            Polat, {\.I}lhan and Feng, Yu and Moore, Eric W. and
            {VanderPlas}, Jake and Laxalde, Denis and Perktold, Josef and
            Cimrman, Robert and Henriksen, Ian and Quintero, E. A. and
            Harris, Charles R. and Archibald, Anne M. and
            Ribeiro, Ant{\^o}nio H. and Pedregosa, Fabian and
            {van Mulbregt}, Paul and {SciPy 1.0 Contributors}},
  title   = {{{SciPy} 1.0: Fundamental Algorithms for Scientific
            Computing in Python}},
  journal = {Nature Methods},
  year    = {2020},
  volume  = {17},
  pages   = {261--272},
}

@article{breton2025,
    author = {Bret\'{o}n-Fuertes, Elisa and Clemente-Marcuello, Clara and Sanz-Arqu\'{e}, Ver\'{o}nica and Tom\'{a}s-Delgado, Gabriela and Lamata-Ot\'{i}n, Santiago and P\'{e}rez-Mart\'{i}nez, Hugo and G\'{o}mez-Garde\~{n}es, Jes\'{u}s},
    title = {Explosive adoption of corrupt behaviors in social systems with higher-order interactions},
    journal = {Chaos: An Interdisciplinary Journal of Nonlinear Science},
    volume = {35},
    number = {9},
    eid = {091103},
    year = {2025}
}

@article{lee2023signed,
  title={Threshold cascade dynamics on signed random networks},
  author={Lee, Kyu-Min and Lee, Sungmin and Min, Byungjoon and Goh, K.-I.},
  journal={Chaos, Solitons \& Fractals},
  volume={168},
  eid={113118},
  year={2023}
}

@article{chang2011performance,
  title={Performance and reliability of electrical power grids under cascading failures},
  author={Chang, Liang and Wu, Zhigang},
  journal={International Journal of Electrical Power \& Energy Systems},
  volume={33},
  number={8},
  pages={1410--1419},
  year={2011}
}

@article{elliott2014financial,
  title={Financial networks and contagion},
  author={Elliott, Matthew and Golub, Benjamin and Jackson, Matthew O.},
  journal={American Economic Review},
  volume={104},
  number={10},
  pages={3115--3153},
  year={2014}
}

@article{fosdick2018,
  title={Configuring random graph models with fixed degree sequences},
  author={Fosdick, B. K. and Larremore, D. B. and Nishimura, J. and Ugander, J.},
  journal={SIAM Review},
  volume={60},
  number={2},
  pages={315--355},
  year={2018},
  publisher={SIAM}
}

@article{starnini2025,
  title={Opinion dynamics: {S}tatistical physics and beyond},
  author={Michele Starnini and Fabian Baumann and Tobias Galla and David Garcia and Gerardo I{\~n}iguez and M\'{a}rton Karsai and Jan Lorenz and Katarzyna Sznajd-Weron},
  journal={Reviews of Modern Physics},
  volume = {98},
  eid = {035004},
  year={2026}
}

@article{caldarelli2025,
  title        = {The Physics of News, Rumors, and Opinions},
  author       = {Caldarelli, Guido and Artime, Oriol and Fischetti, Giulia and Guarino, Stefano and Nowak, Andrzej and Saracco, Fabio and Holme, Petter and De Domenico, Manlio},
  journal      = {Physics Reports},
  volume = {1186},
  pages = {1--75},
  year         = {2026}
}

@article{battiston2025,
  title={Collective dynamics on higher-order networks},
  author={Federico Battiston and Christian Bick and Maxime Lucas and Ana P. Mill\'{a}n and Per Sebastian Skardal and Yuanzhao Zhang},
  journal={Nature Reviews Physics},
  volume = {8},
  pages = {146--159},
  year={2026}
}

@inproceedings{kkt2003,
 author = {Kempe, D. and Kleinberg, J. and Tardos, \'{E}.},
 title = {Maximizing the spread of influence through a social network},
 booktitle = {Proceedings of the Ninth ACM SIGKDD International Conference on Knowledge Discovery and Data Mining},
 series = {KDD '03},
 year = {2003},
 location = {Washington, DC, USA},
 pages = {137--146},
 publisher = {Association for Computing Machinery},
 address = {New York, NY, USA}
}

@book{valente-book,
  title = {Network Models of the Diffusion of Innovations},
  author = {T. W. Valente},
  year = {1995},
  publisher = {Hampton Press},
  address = {Cresskill, NJ, USA}
}

@ARTICLE{granovetter1978,
  title = {Threshold models of collective behavior},
  author = {M. Granovetter},
  journal = {American Journal of Sociology},
  year = {1978},
  volume = {83},
  number = {6},
  pages = {1420--1443},
  }

@article{bootstrap1979,
  author={J. Chalupa and P. L. Leath and G. R. Reich},
  title={Bootstrap percolation on a {B}ethe lattice},
  journal={Journal of Physics C: Solid State Physics},
  volume={12},
  number={1},
  pages={L31--L35},
  year={1979}
}

@article{brummitt2012multiplexity,
  title={Multiplexity-facilitated cascades in networks},
  author={Brummitt, Charles D. and Lee, Kyu-Min and Goh, K.-I.},
  journal={Physical Review E},
  volume={85},
  number={4},
  eid={045102},
  year={2012}
}

@article{hebert2010propagation,
  title={Propagation dynamics on networks featuring complex topologies},
  author={H{\'e}bert-Dufresne, Laurent and No{\"e}l, Pierre-Andr{\'e} and Marceau, Vincent and Allard, Antoine and Dub{\'e}, Louis J.},
  journal={Physical Review E},
  volume={82},
  number={3},
  eid={036115},
  year={2010}
}

@article{schafer2018dynamically,
  title={Dynamically induced cascading failures in power grids},
  author={Sch{\"a}fer, Benjamin and Witthaut, Dirk and Timme, Marc and Latora, Vito},
  journal={Nature Communications},
  volume={9},
  number={1},
  eid={1975},
  year={2018}
}

@article{gleeson2013systemic,
  title={Systemic risk in banking networks without {Monte Carlo} simulation},
  author={Gleeson, James P. and Hurd, T. R. and Melnik, Sergey and Hackett, Adam},
  journal={Advances in Network Analysis and its Applications},
  pages={27--56},
  year={2013}
}

@article{neuhauser2020multibody,
  title={Multibody interactions and nonlinear consensus dynamics on networked systems},
  author={Neuh{\"a}user, Leonie and Mellor, Andrew and Lambiotte, Renaud},
  journal={Physical Review E},
  volume={101},
  number={3},
  eid={032310},
  year={2020}
}

@article{hurd2013watts,
  title={On {W}atts' cascade model with random link weights},
  author={Hurd, Thomas R. and Gleeson, James P.},
  journal={Journal of Complex Networks},
  volume={1},
  number={1},
  pages={25--43},
  year={2013}
}

@article{karimi2013threshold,
  title={Threshold model of cascades in empirical temporal networks},
  author={Karimi, Fariba and Holme, Petter},
  journal={Physica A: Statistical Mechanics and its Applications},
  volume={392},
  number={16},
  pages={3476--3483},
  year={2013}
}

@article{sampson2024,
  title = {Oscillatory and excitable dynamics in an opinion model with group opinions},
  author = {Sampson, Corbit R. and Restrepo, Juan G. and Porter, Mason A.},
  journal = {Physical Review E},
  volume = {112},
  issue = {2},
  eid = {024303},
  year = {2025}
}

@article{hickok2022,
  title={A bounded-confidence model of opinion dynamics on hypergraphs},
  author={Hickok, Abigail and Kureh, Yacoub and Brooks, Heather Z. and Feng, Michelle and Porter, Mason A.},
  journal={SIAM Journal on Applied Dynamical Systems},
  volume={21},
  number={1},
  pages={1--32},
  year={2022}
}

@article{schawe2022,
  title={Higher order interactions destroy phase transitions in {D}effuant opinion dynamics model},
  author={Schawe, Hendrik and Hern{\'a}ndez, Laura},
  journal={Communications Physics},
  volume={5},
  number={1},
  eid={32},
  year={2022}
}

@article{bick2023,
author = {Bick, Christian and Gross, Elizabeth and Harrington, Heather A. and Schaub, Michael T.},
title = {What are higher-order networks?},
journal = {SIAM Review},
volume = {65},
number = {3},
pages = {686--731},
year = {2023}
}

@article{melnik2013,
    author = {Melnik, Sergey and Ward, Jonathan A. and Gleeson, James P. and Porter, Mason A.},
    title = {Multi-stage complex contagions},
    journal = {Chaos: An Interdisciplinary Journal of Nonlinear Science},
    volume = {23},
    eid = {013124},
    year = {2013}
}

@book{lehmann2018,
  title={Complex Spreading Phenomena in Social Systems: Influence and Contagion in Real-World Social Networks},
  editor={Sune Lehmann and Yong-Yeol Ahn},
  year={2018},
  publisher={Springer},
  address = {Cham, Switzerland}
}

@article{kuehn2020,
  title={Adaptive voter model on simplicial complexes},
  author={Horstmeyer, L. and Kuehn, C.},
  journal={Physical Review E},
  volume={101},
  eid={022305},
  year={2020}
}

@article{battiston2020networks,
  title={Networks beyond pairwise interactions: {S}tructure and dynamics},
  author={Battiston, Federico and Cencetti, Giulia and Iacopini, Iacopo and Latora, Vito and Lucas, Maxime and Patania, Alice and Young, Jean-Gabriel and Petri, Giovanni},
  journal={Physics Reports},
  volume={874},
  pages={1--92},
  year={2020}
}

@article{battiston2021physics,
  title={The physics of higher-order interactions in complex systems},
  author={Battiston, Federico and Amico, Enrico and Barrat, Alain and Bianconi, Ginestra and Ferraz de Arruda, Guilherme and Franceschiello, Benedetta and Iacopini, Iacopo and K{\'e}fi, Sonia and Latora, Vito and Moreno, Yamir and others},
  journal={Nature Physics},
  volume={17},
  number={10},
  pages={1093--1098},
  year={2021}
}

@article{benson2018simplicial,
  title={Simplicial closure and higher-order link prediction},
  author={Benson, Austin R. and Abebe, Rediet and Schaub, Michael T. and Jadbabaie, Ali and Kleinberg, Jon},
  journal={Proceedings of the National Academy of Sciences of the United States of America},
  volume={115},
  number={48},
  pages={E11221--E11230},
  year={2018},
  publisher={National Acad Sciences}
}

@book{bianconi2021higher,
  title={Higher-Order Networks},
  author={Bianconi, Ginestra},
  year={2021},
  publisher={Cambridge University Press},
  address = {Cambridge, UK}
}

@article{burgio2023adaptive,
  title={Characteristic scales and adaptation in higher-order contagions},
  author={Burgio, Giulio and St-Onge, Guillaume and H{\'e}bert-Dufresne, Laurent},
  journal={Nature Communications},
  volume = {16},
  eid = {4589},
  year={2025}
}

@article{carletti2020dynamical,
  title={Dynamical systems on hypergraphs},
  author={Carletti, Timoteo and Fanelli, Duccio and Nicoletti, Sara},
  journal={Journal of Physics: Complexity},
  volume={1},
  number={3},
  eid={035006},
  year={2020}
}

@article{centola2010spread,
  title={The spread of behavior in an online social network experiment},
  author={Centola, Damon},
  journal={Science},
  volume={329},
  number={5996},
  pages={1194--1197},
  year={2010}
}

@article{chen2025simple,
  title={A simple model of global cascades on random hypergraphs},
  author={Chen, Lei and Zhu, Yanpeng and Zhu, Jiadong and Cui, Longqing and Ruan, Zhongyuan and Small, Michael and Christensen, Kim and Liu, Run-Ran and Meng, Fanyuan},
  journal={Chaos, Solitons \& Fractals},
  volume={193},
  eid={116081},
  year={2025}
}

@article{ferraz2024contagion,
  title={Contagion dynamics on higher-order networks},
  author={Ferraz de Arruda, Guilherme and Aleta, Alberto and Moreno, Yamir},
  journal={Nature Reviews Physics},
  volume={6},
  number={8},
  pages={468--482},
  year={2024}
}

@article{gleeson2007seed,
  title={Seed size strongly affects cascades on random networks},
  author={Gleeson, James P. and Cahalane, Diarmuid J.},
  journal={Physical Review E},
  volume={75},
  number={5},
  eid={056103},
  year={2007}
}

@article{gleeson2011high,
  title={High-accuracy approximation of binary-state dynamics on networks},
  author={Gleeson, James P.},
  journal={Physical Review Letters},
  volume={107},
  number={6},
  eid={068701},
  year={2011}
}

@article{gleeson2013,
  title = {Binary-State Dynamics on Complex Networks: {P}air Approximation and Beyond},
  author = {Gleeson, James P.},
  journal = {Physical Review X},
  volume = {3},
  issue = {2},
  eid = {021004},
  year = {2013}
}

@article{iacopini2019simplicial,
  title={Simplicial models of social contagion},
  author={Iacopini, Iacopo and Petri, Giovanni and Barrat, Alain and Latora, Vito},
  journal={Nature Communications},
  volume={10},
  eid={2485},
  year={2019}
}

@article{juul2019,
  title={Hipsters on networks: {H}ow a minority group of individuals can lead to an antiestablishment majority},
  author={Juul, J. S. and Porter, M. A.},
  journal={Physical Review E},
  volume={99},
  number={2},
  eid={022313},
  year={2019}
}

@article{kim2023contagion,
  title={Contagion dynamics on hypergraphs with nested hyperedges},
  author={Kim, Jihye and Lee, Deok-Sun and Goh, K.-I.},
  journal={Physical Review E},
  volume={108},
  number={3},
  eid={034313},
  year={2023}
}

@article{kim2024competition,
  title={Competition between group interactions and nonlinearity in voter dynamics on hypergraphs},
  author={Kim, Jihye and Lee, Deok-Sun and Min, Byungjoon and Porter, Mason A. and Miguel, Maxi San and Goh, K.-I.},
  journal={Physical Review E},
  volume = {111},
  eid = {L052301},
  year={2025}
}

@article{landry2020effect,
  title={The effect of heterogeneity on hypergraph contagion models},
  author={Landry, Nicholas W. and Restrepo, Juan G.},
  journal={Chaos: An Interdisciplinary Journal of Nonlinear Science},
  volume={30},
  eid = {103117},
  year={2020}
}

@article{Landry_XGI_2023,
author = {Landry, Nicholas W. and Lucas, Maxime and Iacopini, Iacopo and Petri, Giovanni and Schwarze, Alice and Patania, Alice and Torres, Leo},
title = {{XGI: {A} {P}ython package for higher-order interaction networks}},
journal = {Journal of Open Source Software},
publisher = {The Open Journal},
year = {2023},
volume = {8},
number = {85},
eid = {5162}
}

@article{majhi2022dynamics,
  title={Dynamics on higher-order networks: {A} review},
  author={Majhi, Soumen and Perc, Matja{\v{z}} and Ghosh, Dibakar},
  journal={Journal of the Royal Society Interface},
  volume={19},
  number={188},
  eid={20220043},
  year={2022}
}

@article{sprague2017evidence,
  title={Evidence for complex contagion models of social contagion from observational data},
  author={Sprague, Daniel A. and House, Thomas},
  journal={PloS ONE},
  volume={12},
  number={7},
  eid={e0180802},
  year={2017}
}

@article{stehle2011high,
  title={High-resolution measurements of face-to-face contact patterns in a primary school},
  author={Stehl{\'e}, Juliette and Voirin, Nicolas and Barrat, Alain and Cattuto, Ciro and Isella, Lorenzo and Pinton, Jean-Fran{\c{c}}ois and Quaggiotto, Marco and Van den Broeck, Wouter and R{\'e}gis, Corinne and Lina, Bruno and Vanhems, Philippe},
  journal={PloS ONE},
  volume={6},
  number={8},
  eid={e23176},
  year={2011}
}

@article{st2021master,
  title={Master equation analysis of mesoscopic localization in contagion dynamics on higher-order networks},
  author={St-Onge, Guillaume and Thibeault, Vincent and Allard, Antoine and Dub{\'e}, Louis J. and H{\'e}bert-Dufresne, Laurent},
  journal={Physical Review E},
  volume={103},
  number={3},
  eid={032301},
  year={2021}
}

@article{stonge2022,
  title={Influential groups for seeding and sustaining nonlinear contagion in heterogeneous hypergraphs},
  author={St-Onge, Guillaume and Iacopini, Iacopo and Latora, Vito and Barrat, Alain and Petri, Giovanni and Allard, Antoine and H{\'e}bert-Dufresne, Laurent},
  journal={Communications Physics},
  volume={5},
  number={1},
  eid={25},
  year={2022}
}

@article{watts2002simple,
  title={A simple model of global cascades on random networks},
  author={Watts, Duncan J.},
  journal={Proceedings of the National Academy of Sciences of the United States of America},
  volume={99},
  number={9},
  pages={5766--5771},
  year={2002}
}

@article{zhang2023higher,
  title={Higher-order interactions shape collective dynamics differently in hypergraphs and simplicial complexes},
  author={Zhang, Yuanzhao and Lucas, Maxime and Battiston, Federico},
  journal={Nature Communications},
  volume={14},
  number={1},
  eid={1605},
  year={2023}
}

@article{peralta2020binary,
  title={Binary-state dynamics on complex networks: {S}tochastic pair approximation and beyond},
  author={Peralta, Antonio F. and Toral, Ra{\'u}l},
  journal={Physical Review Research},
  volume={2},
  number={4},
  eid={043370},
  year={2020}
}


\end{document}